
\input harvmac

\def \vx {\vec x}
\def \vy {\vec y}
\def\ma{\mapsto}

\def \up {\uparrow}

\def \pa {\Vert}
\def\ma{\mapsto}

\def\vp{{\bf p}}

\def \ep{\epsilon}

\def \F {{\cal F}}

\def \del {\partial}

\def \ha{{\textstyle{1\over 2}}}

\def \a {\alpha}

\def \F {{\cal F}}\def \B {{\cal B}}\def \t {\theta}\def \tt {\tilde
\theta}

\def \H {\td H} 

\def \s {\sigma}
\def \p {\phi}
\def \m {\mu}

\def \vp {\varphi }

\def \t {\theta}
\def \td {\tilde }

\def \inv {^{-1}}
\def \ov {\over }
\def \four{{\textstyle{1\over 4}}}

\def \C {{\cal C}}

\def \sins{{\rm sin}^2}
\def \coss{{\rm cos}^2}
\def \pab {{\hat \pa}}
\def \botb {{\hat \bot}}

\def \lr { \lref}
\def\np {{  Nucl. Phys. }}
\def \pl {{  Phys. Lett. }}
\def \mpl {{ Mod. Phys. Lett. }}
\def \prl {{  Phys. Rev. Lett. }}
\def \pr  {{ Phys. Rev. }}

\def \cqg {{ Class. Quant. Grav. }}

\baselineskip8pt
\Title{
\vbox
{\baselineskip 6pt{\hbox{   }}{\hbox
{Imperial/TP/96-97/25}}{\hbox{hep-th/9702163}} {\hbox{
  }}} }
{\vbox{\centerline {Composite  BPS configurations 
of p-branes  } 
\vskip4pt
 \centerline { in 10 and 11 dimensions} 
}}
\vskip -32 true pt
\bigskip
\centerline{   A.A. Tseytlin\footnote{$^{\star}$}{\baselineskip8pt
e-mail address: tseytlin@ic.ac.uk}\footnote{$^{\dagger}$}
{\baselineskip8pt Also at  Lebedev  Physics
Institute, Moscow.}}

\smallskip
 \centerline {\it  Theoretical Physics Group, Blackett Laboratory,}
\smallskip
\centerline {\it  Imperial College,  London SW7 2BZ, U.K. }

\bigskip
\centerline {\bf Abstract}
\medskip
\medskip
\baselineskip10pt
\noindent
We  give an overview  of  various composite  BPS configurations 
of string theory and M-theory p-branes  represented as classical 
supergravity solutions. 
Type II string backgrounds can be obtained  by  
S- and T- dualities from  the NS-NS configurations corresponding to exact 
conformal sigma models. The single-center solutions 
 can be also generated from the Schwarzschild solution
by  applying a sequence of boosts, `smearings' in some number of dimensions, dualities
and  taking the extremal limit. 
The basic `marginal'  backgrounds 
representing   threshold BPS  bound states of branes
are parametrised by a number of independent harmonic functions, one 
for each brane.
`Non-marginal' BPS configurations in $D=10$ can be constructed from 
  the marginal ones    using   
 U-duality and thus are parametrised,  in addition to harmonic functions, 
 by a 
finite number of  $O(d,d)$ and $SL(2,R)$  `angles'.  
  Some of them can be viewed  as 
dimensional reductions of 
 coordinate-transformed (boosted
or rotated)  marginal configurations of M-branes. 
We  present   a new  more general class of configurations 
in which some of the branes or their intersection spaces   are localised on 
other branes. In particular, we  find  the  supergravity background describing 
the  type II  BPS configuration of a 3-brane, RR 5-brane and NS-NS 5-brane, 
and   related    `localised' 2-5-5  $D=11$ solution.
We  also consider   the classical action for a 3-brane probe  moving 
in such  type IIB backgrounds and determine the structure of 
the corresponding moduli space metrics.

\bigskip
\vskip 30 true pt


\Date {February  1997}
\noblackbox
\baselineskip 14pt plus 2pt minus 2pt

\lr \dgh {A. Dabholkar, G.W. Gibbons, J. Harvey and F. Ruiz Ruiz,  \np
B340 (1990) 33.
}
\lr\mon{J.P. Gauntlett, J.A. Harvey and J.T. Liu, \np B409 (1993) 363.}
\lr\chs{C.G. Callan, J.A. Harvey and A. Strominger, 
\np { B359 } (1991)  611.}
\lr \CT{M. Cveti\v c and  A.A.  Tseytlin, 
\pl { B366} (1996) 95, hep-th/9510097;
 \pr D53   (1996)  5619, hep-th/9512031.
}
\lr \CTT{M. Cveti\v c and  A.A.  Tseytlin, 
\pr D53   (1996)  5619, hep-th/9512031. 
}
\lr\TT{A.A. Tseytlin, \mpl A11 (1996) 689,   hep-th/9601177.}
\lr \HT{ G.T. Horowitz and A.A. Tseytlin,  \pr { D51} (1995) 
2896, hep-th/9409021.}
\lr\khu{R. Khuri, \np B387 (1992) 315; \pl B294 (1992) 325.}

\lr\ght{G.W. Gibbons, G.T. Horowitz and P.K. Townsend, \cqg 12 (1995) 297,
hep-th/9410073.}
\lr\dul{M.J. Duff and J.X. Lu, \np B416 (1994) 301, hep-th/9306052. }
\lr\hst {G.T. Horowitz and A. Strominger, hep-th/9602051.}
\lr\dull{M.J. Duff and J.X. Lu, \pl B273 (1991) 409. }
\lr \guv{R. G\"uven, \pl B276 (1992) 49. }
\lr \dust { M.J. Duff and  K.S. Stelle, \pl B253 (1991) 113.}

\lr\hos{G.T.~Horowitz and A.~Strominger, Nucl. Phys. { B360}
(1991) 197.}
\lr\gibb{G.W. Gibbons and P.K. Townsend, \prl  71
(1993) 3754, hep-th/9307049.}
\lr\town{P.K. Townsend, hep-th/9512062.}
\lr\kap{D. Kaplan and J. Michelson, hep-th/9510053.}
\lr\hult{
C.M. Hull and P.K. Townsend, Nucl. Phys. { B438} (1995) 109.}
\lr \papd{G. Papadopoulos and P.K. Townsend, \pl B380 (1996) 273, hep-th/9603087.}
\lr\jch {J. Polchinski, S. Chaudhuri and C.V. Johnson, 
hep-th/9602052.}

\lr \duflu { M.J. Duff and J.X. Lu, \np B354 (1991) 141. } 
\lr \pol { J. Polchinski, \prl 75 (1995) 4724,  hep-th/9510017.} 
\lr \iz { J.M. Izquierdo, N.D. Lambert, G. Papadopoulos and 
P.K. Townsend,  \np B460 (1996) 560, hep-th/9508177. }

\lr \US{M. Cveti\v c and  A.A.  Tseytlin, 
\pl {B366} (1996) 95, hep-th/9510097;   hep-th/9512031.  
}
\lr \green{M.B. Green and M. Gutperle, hep-th/9604091.}
\lr \berg{E. Bergshoeff, C. Hull and T. Ort\' \i n, \np B451 (1995) 547, hep-th/9504081.}

\lr \gibbon{G.W. Gibbons, \np B207 (1982) 337. }
\lr \hullo{C.M. Hull, \pl B139 (1984) 39. }
\lr \horts{G.T. Horowitz and A.A.  Tseytlin, \pr D51 (1994) 3351,
hep-th/9409021;   \pr {D50} (1995) 5204, hep-th/9406067.}
\lr \horow{  J.H. Horne, G.T. Horowitz and 
 A.R. Steif, \prl 68 (1992) 568. }
\lr \dabwal {A. Dabholkar, J.P. Gauntlett, J.A. Harvey  and
 D. Waldram, \np B474 (1996) 85,  
hep-th/9511053. 
 }
\lr \tset  {A.A.  Tseytlin,  \np B475 (1996) 179, hep-th/9604035.}

\lr \john {J.H.  Schwarz, \pl B360 (1995) 13 (E: B364 (1995) 252),
hep-th/9508143, hep-th/9509148.}

\lr \johnt {J.H.  Schwarz, \pl B367 (1996) 97, 
 hep-th/9510086. } 

\lr \townelev{ P.K. Townsend, \pl B350 (1995) 184, hep-th/9501068.  }
\lr \calmalpeet {C.G. Callan, J.M.  Maldacena and A.W. Peet, 
\np B475 (1996) 645,  hep-th/9510134. }
\lr \klts{I.R. Klebanov and A.A. Tseytlin,
 \np B475 (1996) 179,
hep-th/9604166. }
\lr \cvets{ M. Cveti\v c  and A.A. Tseytlin, \np B478 (1996) 181, 
hep-th/9606033. }
 
\lr\paptkk{
G. Papadopoulos  and P.K. Townsend, hep-th/9609095. }

\lr\papadop{
G. Papadopoulos, hep-th/9604068. }

\lr \witten {E. Witten, \np B460 (1995) 335, hep-th/9510135.}

\lr \tow {P.K. Townsend,  hep-th/9609217. } 

\lr \schwa  {J.H. Schwarz,
hep-th/9607201.  }
\lr \duff { M.J. Duff, hep-th/9608117.  }
\lr \bergsh{ E. Bergshoeff, E.  Sezgin and P.K. Townsend, \pl B189 (1987)
75.}
\lr \doug {M.R. Douglas,  hep-th/9512077.}
\lr \gaunt {J. Gauntlett, D. Kastor and J. Traschen, hep-th/9604189.}
\lr \aspin {P. Aspinwall,  hep-th/9508154.   }

\lr \CM {C.G. Callan and  J.M.  Maldacena,
 \np B472 (1996) 591, hep-th/9602043.}

\lr\grepap{
M.B. Green, N.D. Lambert, G. Papadopoulos  and P.K. Townsend, 
hep-th/9605146. }

\lr\banks{
T. Banks, W. Fischler, S.H. Shenker and L. Susskind, 
   hep-th/9610043.}

\lr \tser{A.A. Tseytlin, \np B487 (1997) 141, 
 hep-th/9609212. }

\lr\costa { M.S. Costa, hep-th/9609181.}

\lr \schmid {C.  Schmidhuber, \np B467 (1996) 146,   hep-th/9601003.  }

\lr \tsetli{ A.A. Tseytlin, \np B469 (1996) 51, hep-th/9602064.  } 

\lr  \witt { E. Witten, \np B443 (1995) 85, hep-th/9510135. } 

 \lr \grg{ M.B. Green and M. Gutperle, hep-th/9612127.}

 \lr\john{J.H. Schwarz, \pl B360 (1995)13, hep-th/9508143.}

\lr\towns{P.K. Townsend, \pl B373 (1996) 68,   hep-th/9512062.}
 \lr\dull{M.J. Duff and J.X. Lu, \np  B390 (1993) 276.}
 
\lr\gil{G. Lifschytz, hep-th/9612223.}
\lr\gilma{G. Lifschytz and S.D. Mathur, hep-th/9612087.}
\lr \lif{G. Lifschytz, hep-th/9610125.}

\lr\mye{J.C. Breckenridge, G. Michaud and R.C. Myers, hep-th/9611174.}
\lr\rust{J. Russo and A.A. Tseytlin,
\np B490 (1997) 121,  hep-th/9611047.}

\lr\tsey{A.A. Tseytlin, \prl  78 (1997) 1864, hep-th/9612164.}

 \lr\dddo{M.R. Douglas, D. Kabat, P. Pouliot and  S.H. Shenker,
\np B485 (1997) 85,   hep-th/9608024. }

\lr\beh {K. Behrndt, E. Bergshoeff and B. Janssen, hep-th/9604168
(revised).}

\lr \ban{ T. Banks, N.  Seiberg  and S. Shenker, hep-th/9612157.}

\lr\tsee{A.A. Tseytlin, hep-th/9612164.}

\lr \englert {R. Argurio, F. Englert and L. Houart, hep-th/9701042.}

\lr\arfiy{I.Ya. Aref'eva and O.A. Rytchkov, hep-th/9612236.}

\lr\rabi{A.  Giveon, M.  Porrati  and  E. Rabinovici, Phys. Repts. 
244 (1994) 77.}
\lr\rust{J.G. Russo and A.A. Tseytlin, hep-th/9611047.}

\lr\bal{ V. Balasubramanian and R.G. Leigh, hep-th/9611165.}

\lr \iz { J.M. Izquierdo, N.D. Lambert, G. Papadopoulos and 
P.K. Townsend,  Nucl. Phys.  B460 (1996) 560, hep-th/9508177. }

\lr\ber{ M. Berkooz, M.R. Douglas  and R.G. Leigh,
hep-th/9606139.}
\lr \papcost{ M.S. Costa and  G. Papadopoulos, 
hep-th/9612204.  }

\lr \gregut{M.B. Green and 
M. Gutperle, \np B476 (1996) 484, hep-th/9604091.}
\lr \gaukas {J. Gauntlett, unpublished.}
\lr \tooo{A.A. Tseytlin, \pl B395 (1997) 24, hep-th/9611111.}
\lr \ggp{ G.W. Gibbons, M.B.  Green
and M.J.  Perry, 
  \pl  B370 (1996) 37, hep-th/9511080.}
\lr\US{ C.G. Callan,  S.S. Gubser, I.R. Klebanov and 
A.A. Tseytlin, \np B489 (1997) 141, hep-th/9610172.}

\lr\HM{G.T. Horowitz and D. Marolf, hep-th/9610171.}

  \lr\dufe{M.J.  Duff, S. Ferrara, R.R. Khuri and J. Rahmfeld, \pl { B356} 
(1995) 479, hep-th/9506057.}

\lr\khu{ R.R. Khuri,  \pr  D48 (1993) 2947.}

\lr\beh{
 K. Behrndt, E. Bergshoeff and  B. Janssen, 
 hep-th/9604168;  
 E. Bergshoeff, M. de Roo, E. Eyras, B. Janssen and  J.P. van
    der Schaar, hep-th/9612095.} 
 
\lr\gaugi{J.P. Gauntlett, G.W. Gibbons, G. Papadopoulos and P.K. Townsend,
hep-th/9702202.}  
\lr \gauk{J.P.~Gauntlett, D.A.~Kastor and J.~Traschen, 
 \np  B478 (1996) 544,
hep-th/9604179.}

\lr\pope  {   }
\lr\stelle { }
\lr\duff { M.J. Duff, R. Khuri  and J.X. Lu, Phys. Rept. 259 (1995) 213, hep-th/9412184; K.S. Stelle, hep-th/9701088;
  H. L\"u and  C.N. Pope,  hep-th/9702086.    }
  
\lr \hanw{ A. Hanany and E. Witten, 
hep-th/9611230.}
\lr\tsek{I.R. Klebanov and A.A. Tseytlin,
 \np B475 (1996) 179,
hep-th/9604166. }
\lr\bal {V. Balasubramanian and F.  Larsen, 
\np B478 (1996) 199, hep-th/9604189.}

\lr \polch {  J. Polchinski, \prl 75 (1995) 4724,  hep-th/9510017. }
\lr \pol{  J. Polchinski,  hep-th/9611050.}

\lr \bakas { C. Bachas, \pl B374 (1996) 37, hep-th/9511045.           }

\lr \doug  { M.R. Douglas, D. Kabat, P. Pouliot and  S.H. Shenker,
  hep-th/9608024.    } 
\lr\me{A.A. Tseytlin, \np B469 (1996) 51,  hep-th/9602064.}
\lr\greee{M.B. Green and M. Gutperle, \pl B377 (1996) 28,        hep-th/9602077.}  
\lr \hankl{A. Hanany and I.R. Klebanov, 
 \np B482 (1996) 105,  hep-th/9606136; 
I.R. Klebanov and A.A. Tseytlin,  
 \np B479 (1996) 319, 
 hep-th/9607107. }

\lr\garf{ D. Garfinkle, \pr D46 (1992) 4286;
D. Garfinkle and T. Vachaspati, \pr D42 (1990) 1960;
N. Kaloper, R.C. Myers and  H. Roussel, 
      hep-th/9612248.} 

\lr\stromi{A. Strominger,  \pl B383 (1996) 44, 
   hep-th/9512059.}

\lr \pope {N. Khvengia, Z. Khvengia, H. L\"u and C.N. Pope,
hep-th/9605077. }

\lr\reviews{J.H. Schwarz,  hep-th/9607201; M.J. Duff, hep-th/9608117;
P.K. Townsend, hep-th/9609217.}

\lr\instan{N. Ishibashi, H. Kawai, Y. Kitazawa and A. Tsuchiya, hep-th/9612115;
A.A. Tseytlin, hep-th/9701125; I. Chepelev, Y. Makeenko and K. Zarembo, 
hep-th/9701151; A. Fayyazuddin and D. J. Smith, hep-th/9701168.}

\lr\karl{U. Lindstr\"om  and M. Ro\v cek, 
 Commun. Math. Phys. 115 (1987) 21.  } 
\newsec{Marginal   and non-marginal
BPS  configurations of branes: an overview}

\subsec{Introduction}

Viewing the   $D=10$ and $D=11$  supergravities as the  low-energy limits
of the superstring theories and M-theory, it is important to 
have  a better   understanding of the structure of 
 the  space of their classical BPS 
  solutions.  Being  supersymmetric, 
such  solutions are expected to 
encode  useful 
information  about the corresponding states of the full  quantum theory.

The existence of classical solutions describing
 BPS configurations of branes indicates 
a possibility  of  existence of the corresponding   quantum bound states. 
The structure  of  actions of  
classical $p$-brane probes  propagating 
in  supergravity 
 backgrounds produced by configurations of other branes 
gives (at least partial) information 
about  related  quantum theories.
T- and S- duality connections  between configurations 
of branes in $D=10$ imply certain relations between 
their counterparts in $D=11$ so that their study may help 
to  identify hidden symmetries of (quantum)  M-theory
which are not explicit in the $D=11$ supergravity action.
Finally, intersections of branes wrapped over internal spaces 
represent lower-dimensional black holes and thus guide the  
studies of the black-hole  properties by suggesting  which 
 configurations of 
`microscopic'  branes admitting  quantum-mechanical description
should be considered.

The stationary  supergravity solutions representing 
composite  BPS configurations of branes 
can be classified into  `marginal' (or `threshold') 
  and  `non-marginal'  ones.
The  marginal backgrounds are the basic ones
while the $D=10$ non-marginal configurations  
fall into  families of descendants  of a `core'  marginal 
solution to which they are related by U-duality.\foot{In what follows U-duality \hult\  will mean a superposition
of  T- (i.e. $O(d,d)$)  and S- (i.e. $SL(2,R)$)  duality transformations.
T-duality will be assumed to  act 
 in all possible  isometric directions, including 
time. We shall  discuss composite configurations of branes in $D=10$ and $11$; 
for reviews of $p$-brane solutions in various  dimensions 
see \refs{\duff}.}

The  marginal solutions are parametrised by a number $N$ 
of 
independent harmonic functions $H_i(x)$   which is equal to 
the number of branes 
in the configuration   (counting also a possible 
wave along null direction).\foot{The  simplest
 example of a composite BPS solution 
parametrised by {\it two}
 independent harmonic functions  is  the  
superposition 
  of the fundamental string \dgh\ and a plane wave   \horts, 
representing, in the 1-center case, 
 a BPS string state with a momentum flow 
along the string \refs{\dabwal,\calmalpeet}.
Another   example  is a superposition of a fundamental string and 
a solitonic 5-brane $1\Vert 5$  \refs{\CT,\TT}.
The existence of such NS-NS sector  solutions parametrised by several  harmonic 
functions
follows directly from  the conformal invariance 
condition on the string sigma-model  (`chiral null model' \horts).
The $1\pa 5$  configuration served as a starting point for the construction 
 of various intersecting \papd\ 
 brane configurations in $D=10,11$ which are parametrised 
by several harmonic functions according to the `harmonic function rule' \tset.}
 In the simplest  case  of `delocalised' combinations  of branes
when all internal dimensions of the branes are isometries
(so that  the configuration can be viewed as an `anisotropic' brane \guv)
the functions $H_i$ satisfy the free flat-space Laplace equation
with respect to the common transverse space coordinates. In 
 more general cases
of `localised'  intersections of branes discussed below
some of $H_i$ may depend on internal coordinates of some branes
and satisfy  curved-space Laplace equations.
When all of the harmonic functions have singularities at  the same center, 
the mass of the marginal  configuration is  proportional to
 the sum of the `charges', \ 
$M= Q_1+...+Q_N$.

Marginal configurations with the same number of (families of parallel) 
branes $N$
(that means, typically,  with the  same amount $1/2^N$ 
of unbroken supersymmetry)  belong to one  universality class  being 
 related by simple discrete T- and S- duality transformations  
combined with an operation of `smearing' (or `delocalisation', or 
forming an infinite  periodic array)
in some number of transverse dimensions.
Starting with a marginal configuration,
 one can also apply a sequence of 
T- and S- duality transformations
with  arbitrary { continuous} (at the classical level) 
 parameters. 
 The result is a {\it non-marginal}  configuration of branes
which  is  parametrised by $N$  
harmonic functions of its `parent' marginal solution {\it and} 
a finite number of U-duality parameters (angles and boosts of 
$O(d,d)$ duality \rabi\  and entries of the $SL(2,R)$ matrix
 of S-duality \john). Since T- and S-dualities preserve 
supersymmetry, these 
 non-marginal solutions
 have the same amount of unbroken supersymmetry as their `parent'.
From  the lower-dimensional  point of view,  they  correspond
 to U-dual versions  of  black holes obtained by wrapping
 all isometric internal coordinates of a composite 
$p$-brane configuration over a torus.
In the case of single-center harmonic functions
 the non-marginal solution 
will  represent a  configuration with $\td N > N $ charges
and  its  mass  will be  typically of the form $M=
\sum \sqrt {\td Q_{1}^2 + ...+ \td Q^2_{\td N}}$, 
indicating a non-vanishing `binding' energy.

The marginal configurations in $D=10$ have direct counterparts in $D=11$.
One way to construct non-marginal configurations of branes in $D=11$ is to lift up
the $D=10$ 
non-marginal  solutions   found by U-duality from the marginal type II 
theory ones. 
Some of them turn out to be  just 
rotations and finite boosts of marginal M-brane intersections \rust, 
but there are also other non-marginal 
$D=11$ solutions (e.g., the $2+5$ combination  \iz).
The general the rule of constructing  non-marginal $D=11$ configurations 
from marginal ones (i.e. a counterpart of U-duality in $D=10$)
which  applies directly in $D=11$ (i.e.  is 
a certain transformation of the $D=11$ metric and 3-index tensor)
  remains to be  explicitly formulated.
While the S-duality acts in $D=11$ simply 
as a coordinate  transformation `mixing' the directions of 
dimensional reduction and T-duality  from IIA to IIB theory
\refs{\berg,\johnt}, 
the lift to $D=11$ 
of the action of $O(d,d)$ duality
on the space of $D=10$ solutions 
was  not yet  described  in general.
 This  may give an important hint about
 new symmetries of M-theory.

In Section  1.2 we shall  discuss the  construction 
of various composite  BPS configurations 
of type II and M-theory  $p$-branes  represented as classical 
supergravity solutions.  Some type IIB solutions (in particular, 
a  non-marginal combination of a fundamental string 
and an instanton)
will be  described  in Section 1.3.
In Section 2  we shall consider  in more detail the non-marginal 
configurations of branes in $D=10$ and $D=11$, and present some new 
examples of such solutions. 

Section 3 will be devoted to 
 a   more general  class of  `localised' configurations 
which have less isometries than  `smeared' intersections 
of branes  found   previously in \refs{\papd,\tset,\gauk}.
We shall show how they can be constructed by applying 
dualities to the `fundamental string plus  5-brane' type 
NS-NS backgrounds corresponding to exact 
conformal  `chiral null  models'.
 In particular, we  shall find  
  a supergravity background representing 
a type IIB  intersection  of a 3-brane, RR 5-brane and NS-NS 5-brane
(the existence of such  BPS  configuration was pointed out in \hanw).
We shall also 
 discuss  some related solutions  in $D=10$ and $D=11$, e.g., 
a localised configuration of a 2-brane and two 5-branes in $D=11$.
In Section 4  we shall 
 study   the classical actions for  $p$-brane probes  moving 
in such  composite  BPS  backgrounds.

\subsec{Construction of solutions and examples}

The BPS configurations with single-center harmonic functions 
have `non-extremal' generalisations 
(corresponding, upon compactification of isometric internal dimensions,
 to U-duality families of 
non-extremal black holes).
It is remarkable that to  construct such solutions 
there is no need to solve the classical equations explicitly \tooo: 
 all one needs to know  is \    (i) the vacuum  Schwarzschild solution,
 and (ii) T- and S-duality transformation rules  
of  type II supergravity fields \berg. 
The one-center BPS solutions can then be obtained by taking the extremal limit.
Indeed, starting with the neutral black string (i.e. `Schwarzschild $\times R_y$'),
boosting it along isometric $y$-direction 
 and applying T-duality one finds the non-extremal 
version of the fundamental string background \horow.\foot{Infinite boost combined with sending 
the mass of the Schwarzschild solution $\m$ to zero gives a
  plane wave background \gibbon\ which is T-dual to the extremal fundamental string.}
 S-duality relates it to the 
RR string of IIB theory. Adding extra isometries (i.e. smearing in transverse directions) and applying T-duality leads to all other $p>1$ 
 RR $p$-branes \hos. Acting by  S-duality  on the RR 5-brane one 
finds the non-extremal version of NS-NS 5-brane of type IIB theory, which,  in turn, is 
 related by T-duality in longitudinal direction to the (identical) 
5-brane background of type IIA theory.  
More general U-duality transformations lead to 
non-extremal versions of non-marginal BPS configurations 
with 1/2 of supersymmetry.

To find composite brane  solutions  with two charges which 
become marginal 1/4 supersymmetric configurations in the extremal limit
one may start with a black fundamental string 
($ds^2_{10} = H\inv (r) [ - f(r) dt^2 + dy^2] +f\inv (r)  dr^2  + r^2 d\Omega_7^2$, \ $f = 1 - \mu/r^6$, etc.)
and apply a boost. In the limit of the infinite boost  and $\m\to 0$ 
this gives a superposition  of a fundamental string with a wave or  $1\up$.\foot{We shall use the following notation for the bound states of branes.
$p\pa p'$ will denote a marginal composition of  a $p$-brane and $p'$-brane 
in which their internal dimensions are parallel (in the single-center case 
one brane is  on top of the other; 
in the case of multi-center harmonic functions
this is a collection of parallel branes localised at different points).
$p\bot p'$ will denote  a  marginal configuration 
representing orthogonal intersection of two branes  
(they may share some number $q(<p,p'$) of spatial dimensions
and can be separated in transverse dimensions in the case of multi-center 
choice of the two independent harmonic 
functions).	
For  most of the solutions discussed in Sections 1 and 2 
 the configuration of several 
 branes will be  `delocalised'
in all internal dimensions. 
$p\up$ will stand for a  1/4 supersymmetric bound state of a $p$-brane
(with $p >1$)  and a plane wave, i.e. a 
configuration of a $p$-brane with a momentum flow along one of its longitudinal directions (it can be obtained, e.g.,  as  a limit of 
 a non-extremal $p$-brane
infinitely boosted in the longitudinal direction). 
$p+p'$ will denote a non-marginal configuration 
representing a bound state of two branes which cannot be separated 
in the transverse dimensions. This is an  interpolation
 between  a $p$-brane  and a $p'$-brane 
configurations and is parametrised by a single harmonic function
specifying the common center(s) of the two branes forming the
 bound state. $(...)_n$ will indicate that the configuration is smeared over $n$ 
 transverse directions, i.e. 
 has  $n$  extra  isometries.
All  $D=10$ metrics below are the string-frame metrics. 
}

 Similar more general extremal solution is parametrised 
by two independent harmonic functions $H_1$ and $H_w=1+K$
\eqn\sol{ ds^2_{10} = H\inv_1 (x) [ -  dt^2 + dz^2  +  K(x)  (dt-dz)^2 ]
 +  dx_mdx_m  \ , 
}
$$ e^{2\p} =  H_1\inv\ , \ \ \ \ 
\ \ \ 
dB = dH_1\inv \wedge dt \wedge dz \ . $$
The existence of this solution follows from 
conformal invariance of the corresponding `chiral null model' \horts.
This configuration 
serves as a  starting point for the  construction 
of other  marginal  and non-marginal 1/4 supersymmetric
 compositions of branes. 
The $SL(2,R)$ duality converts NS-NS objects  into the RR ones 
and T-duality relates all of the RR $p$-branes.
For example, we get the following U-duality sequence of solutions
(adding  extra transverse isometries by `smearing' the string solution):
$1_{NS}\up\ \ \to\ 1_{R}\up\ \ \to\ 5_{R}\up\ \ \to\  5_{NS}\up\  \ \to\
  5_{NS}\pa 1_{NS}$\
$\to\ 5_{R}\pa 1_{R}\ \to\ 3\bot 3 \to 4+0$, etc.
Another sequence is $1_{NS}\up\ \  \to 1_{R}\up\ \   \to 1\pa 0$, etc. 

In particular, the 
    1/4 supersymmetric  NS-NS background corresponding to
 $  5_{NS}\pa 1_{NS}$
can be  obtained  from the  above   one  $1\up$ \ \sol\ 
just by using the  standard T- and S- duality transformation rules.
This   solution  describing a fundamental string 
smeared over a solitonic 5-brane $5_{NS}\pa 1_{NS}$
was originally  found  directly,  by 
using conformal sigma model considerations 
(which also imply its exactness to all orders in $\a'$) \TT\
\eqn\fio{
ds^2_{10} = H_5(x) [ H_1\inv (x)H_5\inv (x) (-dt^2 + dz^2 ) + 
 H\inv_5(x) dy_n dy_n  + dx_m dx_m ] \ , } 
$$ e^{2\p} = H_5 H_1\inv\ , \ \ \ 
\ \ \ 
dB = dH_1\inv \wedge dt \wedge dz + * dH_5 \ , $$
where $(z,y_n)$  are the internal coordinates of the 5-brane ($n,m=1,2,3,4)$.

To construct   marginal configurations with more than two  charges 
one is to start again with non-extremal solution, 
add extra isometries,  apply boost and T-duality, and take the extremal limit.
In this way one finds the explicit form of the  3-charge configurations like
$5\pa 1\up$, $3\bot 3\bot 3$, etc.
This procedure (i.e. U-duality) 
automatically determines the rules of 
 intersections  of branes which 
are consistent with the marginal  BPS property.
It also  implies the `harmonic function rule'
dictating the dependence of   background fields
on harmonic functions.
Since the  U-duality transformations preserve supersymmetry,
all $p$-branes dual to the  fundamental string have 1/2 of supersymmetry, 
all BPS combinations of branes dual to $1\up$ have 1/4 of supersymmetry and 
configurations dual to $1\pa 5 \up$ have 1/8 of supersymmetry. One is also guaranteed to have the same amount of supersymmetry for the $D=11$ solutions 
obtained by `lifting up'  the type IIA backgrounds.

An  alternative approach to determining the rules of constructing 
marginal configurations of $p$-branes (i.e. the intersection rule and the harmonic function rule) 
which  applies to all possible $p$-brane choices in $D=10$ 
and in $D=11$ is based on consideration of an action of a $p$-brane
probe  moving  in the  supergravity background produced by another $p'$-brane
and imposing the condition
of the vanishing of a 
 force on a static probe (marginal BPS state condition).
This determines the relative orientation of the  $p$-brane probe 
with respect to the $p'$-brane source and thus the intersection rule \tser.
Thus the knowledge of  single-brane solutions  and the basic terms in the actions 
for their collective coordinates  which follow from  the supergravity actions
makes possible to  construct 
 composite configurations of
 branes.\foot{The intersection rules can be found also 
directly  from   the basic  field equations
 \refs{\arfiy,\englert}.}
The conclusion is that  in $D=11$ the following intersections 
represent marginal BPS configurations (as originally suggested 
in  \refs{\papd,\stromi}):
$2\bot 2 (0)$,\  $5\bot 5 (3)$, \ $2\bot 5 (1)$ (figure in brackets is the 
number of common spatial directions of the two orthogonally intersecting
 branes).
In $D=10$ one finds that the following configurations are 
possible: (i) NS-NS intersections: $1 \pa 5$, $5\bot 5 (3)$; (ii) RR intersections:
$p\bot q(n)$, $n= \ha (p+q) -2$, i.e.
$n=0:\  4\pa 0,  3\bot 1, 2\bot 2$;\ $n=1: \ 1\pa 5, 4\bot 2, 3\bot 3$;
$n=2: \ 6\pa 2, 5\bot 3, 4\bot 4$, etc.;
(iii) `mixed'  intersections:  for any RR $p$-brane 
$p_R$ the following intersections are possible:  
$1_{NS} \bot p_R (0)$ and $5_{NS} \bot p_R  (n)$, $n=p-1$.
Examples 
are $1_{NS} \bot 1_R$ (which is T-dual to $(1+0)_1$ or $2\up$)
and    $5_{NS} \bot 1_R (0), \ 5_{NS} \bot 2 (1), \
5_{NS} \bot 5_R (4) $.

The non-marginal solutions are obtained  by applying 
more general T- and S- duality transformations.
 They depend on extra U-duality parameters and thus  
  `interpolate'
between marginal configurations with the same amount of supersymmetry
(and the same number of isometries).
For example, applying $SL(2,R)$ transformation to the fundamental string
$1_{NS}$ one finds the string-string bound state $1_{NS} + 1_R$ \john.
 U-duality then relates this solution to other 1/2 supersymmetric 
non-marginal bound 
states; for example,  
\rust\ \ \ 
$1_{NS} + 1_R \ \to\ 1_{NS} + 3\ \to\ 1_{R} + 3\ \to\ 0 + 2$,
 or \ 
$1_{NS} + 1_R\ \to\ 1_{NS} + 5_R\ \to\ 1_{R} + 5_{NS}\ \to\ 0 + 5.$
The $2+0$ and $5+0$ non-marginal  bound states 
can be obtained   also as  dimensional 
reductions
of the  $D=11$ \  2-brane and 5-brane  finitely boosted in  transverse 
11-th dimension, $2\ma$ and $5\ma$ \rust.
An alternative way   to construct these configurations
is to apply  $O(d,d)$  duality transformations to single $p$-branes
with a number of transverse isometries. 
For example, starting with a  0-brane `smeared' over a line, 
 finitely boosting it along the isometric direction 
and performing T-duality one finds again the  $1_{NS} + 1_R$ solution \rust.
Starting with a RR $p$-brane smeared over one transverse dimension $y$ 
and applying T-duality in the direction rotated  by an angle in the plane formed 
by $y$ and an internal $p$-brane coordinate one finds the 
bound state of a  RR $(p-1)$-brane and a  RR $(p+1)$-brane, for example,
$(1_R)_1\to\ 0+2$ or $(2)_1\to\ 1_R + 3 $ \mye.
More general $O(d,d)$  duality transformations with several 
parameters lead to more complicated 1/2 supersymmetric 
configurations; for example, 
starting with 2-brane with extra 2 or 4 transverse isometries 
one finds  $(2)_2 \to 4+2+ 0$\ \mye\ or $(2)_3 \to 6+4+2+0$.\foot{This 
symbolic notation indicates 
the  nonvanishing charges  present in the non-marginal  bound state
and also branes  which are special cases of this more general 
configuration corresponding to limiting values of the angular parameters
(this notation  ignores the fact that  there are actually 
several 2-brane charges corresponding 
to different orthogonal planes).} 

To find   non-marginal configurations with 1/4 of supersymmetry 
(depending on two harmonic functions and a  number of parameters) 
one may start with
a marginal solution  $1\up$  (or $1+0$,  or  $2\bot 2$, etc.) and 
apply U-duality with arbitrary  `angles' and `boosts'.
This leads, in particular,  to the  explicit form of the 
supergravity background representing the  $(4+2+0)\pa 0$
 configuration  (see Section 2).
Similarly, one can construct families of non-marginal 
configurations  with 1/8 of supersymmetry 
(three harmonic functions) 
by starting with $N>2$  marginal bound states
like $2\bot 2\bot 2$, etc.
They  will include as special cases the non-marginal
 configurations  depending on  $N'<N$  different  harmonic functions
but the same number of U-duality parameters.

Lifting the marginal type IIA solutions to $D=11$ one finds 
the  marginal composite configurations of M-branes:
$2,5, 2\bot 2, 2\bot 5, 2\bot 2\bot 2$, etc. \refs{\papd,\tset,\tsek,\gauk}.
The existence of configurations with longitudinal momentum waves like
$2\up, 5\up, 2\bot 5\up,$ $5\bot 5\up$  
\refs{\tset, \tsek,\pope}
can be viewed as a consequence of the existence of the 
corresponding   $D=10$ solutions in the NS-NS sector, 
$1\up, \ 5\up,\ 1\pa 5\up$  which are 
described by conformal chiral null models
\refs{\horts,\CT,\TT}.  
 These solutions  (and their generalisations to the case when the 
wave harmonic function depends on the null coordinate $u=z-t$) 
can  be also 
constructed  directly in $D=11$ 
following  the approach   of  \garf.

Reducing the $D=11$ solutions to $D=10$ along different directions 
leads to  several  $D=10$ marginal backgrounds  
which thus have a common origin in $D=11$
(for various examples and a classification of such  solutions 
 see \refs{\tset,\tsek,\gauk,\bal,\beh}). 
 In particular,  
$5\up$ and $4+0$ in $D=10$ 
are the reductions of $5\up$ in $D=11$. More complicated example is 
$2\bot 5 $ 
which has the following  counterparts in   $D=10$:
(i) $2\bot 4$  which is T-dual to  $1_{R}\pa 5_{R}$ or 
$3\bot 3$;\ (ii)  $1\bot 4$  which is T-dual to   $5_{R}\up$; \ 
(iii) $1\pa 5$  which is T-dual to  $5_{NS}\up$ and  
$1_{NS}\pa 5_{NS}$. These  T-dual configurations
 are related  by simple S-duality;  
this  is consistent 
with  the $D=11$ interpretation of S-duality of IIB theory 
as 
 a coordinate transformation that  interchanges 
 the directions of dimensional reduction and T-duality 
between  type IIA and  type IIB theories \refs{\berg,\john}.

Thus  part of  U-duality relating different
configurations of  type IIA  branes 
in $D=10$ becomes  simply a coordinate transformation in   $D=11$.
Dimensionally reduced to $D=10$,  the $D=11$  solutions 
with the same $N$ become connected  by U-duality. However, there is no  known 
analogue of U-duality which  would  relate  them 
directly in $D=11$.

 Similar conclusions follow 
 from consideration of non-marginal 
backgrounds. Some of $D=10$ non-marginal solutions
can be viewed as  reductions of coordinate-transformed (rotated or boosted)
marginal configurations in $D=11$ \rust\ (for example, 
as already mentioned above, 
$2+0$  and $5+0$  are reductions of  finitely boosted M-branes).
The coordinate transformations applied to marginal
configurations in $D=11$ do not lead, however, to the full  U-duality 
families of 
non-marginal bound states in $D=10$.
In addition to coordinate transformations
of the marginal configurations 
 there are  also other non-trivial  non-marginal $D=11$ 
 bound states (like $2+5$ \iz\ and its generalisations \costa) 
 which are lifts of other `parts' of U-duality families in $D=10$.
We shall discuss  examples of  new $D=11$ solutions of that kind  in 
Section 2 below.

\subsec{Some  type IIB solutions}
As follows from the above discussion,  all 1/2 supersymmetric
 BPS configurations of $p \geq 0$  branes 
 in $D=10$ can be constructed by 
starting with a 0-brane background, 
smearing it in some number of dimensions and applying U-duality. 
 At the same time, 
the basic object of lowest space-time dimensionality 
is the type IIB instanton  \ggp\
\eqn\inis{ ds^2_{10} = H^{1/2}_{-1} (x)  dx_\m dx_\m\ , \ \ \ \  
 e^\p =  H_{-1} \ , \ \ \ \ \  \C=    (H_{-1})\inv - 1\ 
\  . }
Here $\m=0,1,...,9$ and   $H_{-1}$ is the harmonic function 
($=1 +Q/x^8$ for a single instanton).
 We 
 consider the type IIB theory with
 euclidean time $x_{0} = it$  so that $\C=i C$, where $C$ is the RR scalar. 
Taking a distribution of instantons along  $x_0$, 
so  that $x_0$ becomes an isometry, 
$ ds^2_{10} = H^{1/2}_0  (dx_0^2   + dx_m dx_m ), \ H_0= 1 + Q/x^7$, 
and applying  T-duality in this direction  one finds 
the 0-brane solution \hos\ 
of type IIA theory. The backgrounds for
all  other RR $p$-branes can be constructed in a similar way 
by spreading instantons over a $(p+1)$-dimensional space-time 
and  performing T-duality transformations. 
The NS-NS branes can  then  be obtained  by  S-duality
(for example, 
the  type IIA 
5-brane is  T-S-T dual of  the  instanton 
solution  smeared over a 6-plane).

Starting with a  stationary  type IIA  $p$-brane configuration
and including the
  T-duality along the  isometric  time direction 
 into the set of possible $O(d,d)$ transformations, 
 one is able to construct 
various   `instanton-type' solutions of type IIB theory.
Some of such backgrounds  will have   the
(euclidean)  time
direction  being  orthogonal to the  $p$-brane world-volume  (similar
configurations were considered  in \gregut).

Applying
T-duality in the  time direction  to the  fundamental string background
gives the plane wave background, 
while time-like T-dual  of the 
 $1\pa 0$  bound state  with the metric $ds^2_{10} =
H_0^{1/2}[ - H_0\inv H_1\inv dt^2  + H\inv_1 dz^2 + dx_m dx_m ]$ 
gives a  (complex) 
configuration which  is a `mixture' 
 of an instanton and a plane wave (T-dual of $1\pa 0$ along  the  spatial 
isometry $z$ is $1_R\up$)
\eqn\iss{
ds^2_{10} =
H_0^{1/2}[ -dt^2  +  dz^2 +  (H_1 -1) (dz-dt)^2  + dx_m dx_m ] \ , 
\ \ \  iC = H_0\inv -1  \ , 
\ \ \  e^{\p} = H_0  \ .  }
A non-trivial  example of a marginal 
 euclidean type IIB  configuration  
parametrised by two harmonic functions is  a  bound state of 
a (smeared) instanton and a 3-brane 
$3\Vert (-1)$   which has  
the  metric  \tsee\
\eqn\nst{  ds^2_{10}= (H_{-1}  H_3)^{1/2} 
( H_3^{-1}  dx_k dx_k + dx_mdx_m )\  ,  } 
where $k=1,...,4$,\ $m=5,...,10$, and  $H_{-1} $ and $H_3$ are the 
 harmonic functions depending only on  $x_m$.
This solution is T-dual to $4\pa 0$ or $5_R \pa 1_R$.

Simplest non-marginal   type IIB configurations  
 can be  found by applying  $O(d,d)$  duality
 to a 0-brane with one extra transverse isometry.
Starting with $0\ma $, i.e.  the 
 0-brane finitely boosted in the  isometric direction 
 to the 
 velocity  $v= \cos \t  $   
 \eqn\heq{
ds^2_{10A} =  \td H^{1/2} \big(  -  \td H\inv  d\td t^2 + 
d\td y^2  +  dx_m dx_m \big)  \ , }
$$ dA= d\td H\inv \wedge d\td t \ , \ \ \ \ 
e^{2\p} = \td H^{3/2}  \ , \ \ \ \ \ \td H -1 = \sins \t (H-1) \ , \ \ H=H_0 \ ,  $$
$$ \td t\equiv (\sin \t)\inv  ( t - \cos \t\ y) \ , \ \ \ \ 
\td y \equiv   (\sin \t)\inv  ( y - \cos \t\  t ) 
 \ , 
$$
 and  performing  T-duality in $y$   gives \rust\     
the string-string solution  $1_{NS} + 1_R$ of  \john\ 
\eqn\johhn{
ds^2_{10 B  } =   \H^{1/2} [ H\inv ( -dt^2 + dy^2) + dx_m dx_m ]\ , 
\ \ \ \ \  e^{2\p}=   H\inv  \H^2  \ ,}
\eqn\init{    C =  \sin \t  
 \cos \t\  (H-1)  \H^{-1}  \ ,   \ \ \  \ 
 dB  + i dC_{2}  =   (\cos \t + i\sin \t)\  dH^{-1}\wedge dt\wedge dy \  .  }
T-duality in $t$ direction  produces   the same 
background  with 
 $\cos \t \to {1\ov \cos \t}$, $\sin \t \to {\sin \t \ov \cos \t}$ (note that the boosted configuration
\heq\  is `symmetric'  in $t$ and $y$), i.e. \johhn\ and 
\eqn\ini{C =  { \sin \t \ov \coss \t }  
   (H-1)  \H^{-1}  \ ,  \ \ \  \ 
dB  + i dC_{2}  =    {1+ i \sin \t \ov \cos \t }\ 
  dH^{-1}\wedge dt\wedge dy  \ .  }
If we set $ H -1 = \coss \t (H_{-1} -1)$,  \ 
 $\td H -1 ={ \sins \t \ov \coss \t } (H-1) = \sins \t (H_{-1} -1)$
then this 
non-marginal type IIB 
configuration  can be interpreted as a bound state of  
a  fundamental string  and an  instanton, $1_{NS} + (-1)$.
Indeed, in the zero-boost limit   ($\t={\pi\ov 2}$, \ $H=1,\ \td H=H_{-1}$) 
this becomes the instanton \inis\  (smeared over a 2-plane)
while  in the infinite boost limit 
($\t=0$, \ $H=H_{-1},\ \td H=1$) we get 
 the fundamental string background.\foot{Though the solution  under discussion is a non-marginal one,
it  is interesting to note that the metric \johhn\ 
has the same structure as that of a  would-be marginal  superposition of  an instanton 
and a fundamental string constructed according to the harmonic function rule
with independent functions $\td H$ and $H$.}
  Since both  $1_{NS} + 1_R$ and $1_{NS} + (-1)$
are dual to  $0\ma$
they are  related by $O(2,2)$ duality.

Treating $\t$   as a complex   parameter 
one finds the background which formally 
interpolates between all   three limiting cases: 
$1_{NS}$, $1_R$ and $-1$. 
To clarify why such interpolation is possible,  
let us start with unboosted $0$-brane 
with an extra isometry $y$ and  rotated  time direction  $x_0=i t$.
T-duality along $x_0$  gives the instanton, while T-duality along 
$y$ gives the RR string (continued  to euclidean time). 
T-duality along a rotated direction $y' = \cos \psi\ y + \sin \psi \ x_0$
then produces  the non-marginal $1_R + (-1)$ background.
It has the same structure as  \johhn,\init\ but now with 
$t\to -ix_0$, \ $\td H=H_0, \ H-1= \coss\psi (H_0-1)$
and $\cos \psi = {1\ov \sin\t},\  \sin \psi = i{ \cos \t\ov \sin \t}$ 
($\psi$  plays the role of  an  imaginary boost parameter).
The configuration $1_{NS} + (-1)$  may be of interest in connection 
with instanton matrix model discussed in  \instan.

\newsec{Non-marginal BPS  configurations in $D=10$ and 
in $D=11$}

 To find non-marginal solutions in $D=10$ one 
may  apply  $O(d,d)$  and $SL(2,R)$  dualities to  the marginal configurations.
The basic transformations of $O(d,d)$ duality are `boost + T-duality' 
and `rotation + T-duality'.  
As discussed above, boosting 
a 0-brane in an isometric direction  and applying 
T-duality one finds  $1_{NS} + 1_R$ type IIB configuration.
Starting with the RR string $1_R$  along $y_1$ 
with an extra transverse isometry $y_2$ and  applying T-duality along 
rotated direction in the $(y_1,y_2)$ plane  one obtains  the non-marginal 
$2+0$ 
type IIA configuration.

Below we shall consider  more complicated examples of such 
non-marginal solutions which `interpolate' between their marginal 
limiting cases  and  study  their lifts to $D=11$.
The idea  is to  find $D=11$  solutions which 
are parametrised  by a number of harmonic functions $H_i$ 
and  parameters   $\vp_i$
such that the 
variation of the `angles' connects various  marginal $D=11$ solutions 
with the same 
number of  independent
harmonic functions
and thus the same amount of  supersymmetry.
One way to construct  such   solutions is to start with
 any of  limiting marginal $D=11$ 
backgrounds,  reduce it  down in isometric directions 
 to  some lower $D$, 
apply  U-duality
and then lift the resulting configuration 
 back to $D=11$. 
Studying such general families of solutions may help to 
 learn  how T-duality acts directly in $D=11$:
having   a  generalisation of a simple marginal solution 
`dressed' by U-duality 
 parameters, one may  be able to  extract  the transformation rule of
the  $D=11$ metric and the  $C_3$ field
 that generates it.
 It is not clear a priori 
  whether this procedure leads to  new solutions, 
or, as it happens in $D=10$, they are related to the basic marginal ones by a symmetry.\foot{The  $D=10$
 type II 
supergravity actions are known to be 
 invariant under (or related by)
 T-duality transformations along abelian isometries.
Though T-duality may look  accidental 
  at the level of the supergravity action,
it has a microscopic 
explanation based on  2d duality on the  string world sheet and the fact that 
 $D=10$  supergravity
actions
are the low-energy effective actions of string theories.
When this 
 symmetry is  lifted up to the level of 
$D=11$ supergravity action it becomes 
  truly  miraculous as in $D=11$ 
 there is no  known general  microscopic explanation for it.
}

Let us start with  simplest examples. The fundamental string 
$1_{NS}$  along $y_1$ in a space with one extra isometry 
$y_2$
is transformed by $T$-duality  along  a  rotated direction 
($y'_2= \cos \vp\ y_1 + \sin \vp\ y_2$)
into  $1_{NS}$  finitely boosted  in that angled 
direction (so that  it has 
$H\to \td H =  1 + {\rm sin}^2\vp (H-1)$ 
and $t$ and $y_2'$ `mixed up' with velocity $\cos \vp$). 
Lift to $D=11$ gives a  2-brane finitely boosted 
in an angled direction  with the  second
  internal direction being $y_{11}$
\eqn\ele{
ds^2_{11}= \td H^{1/3} [ \td H^{-1}(-d\td t^2 +  
dy_1^2 + dy^2_{11})
+ d\td y^2_2  + dx_m dx_m ] \ , 
}
$$ \td t = (\sin\vp)\inv( t - \cos \vp\ y_2') \ , 
\ \ \ 
\td y_2 = (\sin\vp)\inv( y_2' - \cos \vp\  t) \ . 
$$
Similarly, applying  T-duality at angle to  $1_R$ \mye\ one finds  $2+0$ 
configuration   
 which  is lifted to a 2-brane boosted along the 
orthogonal $y_{11}$ direction  \rust
 \eqn\elele{
ds^2_{11}= \td H^{1/3} [ \td H^{-1}(-d\td t^2 +  
dy_1^2 + dy^2_2)
+ d\td y^2_{11}  + dx_m dx_m ] \ , 
}
$$ \td t = (\sin\vp)\inv( t - \cos \vp\ y_{11}) \ , 
\ \ \ 
\td y_{11} = (\sin\vp)\inv( y_{11} - \cos \vp\  t) \ . 
$$
These two  backgrounds  correspond
to the type IIB solutions 
 related by  discrete  ($\theta = {\pi \ov 2}$)  $SL(2,R)$ rotation, 
which in  $D=11$ thus  
corresponds to  interchanging the direction of dimensional reduction
$y_{11} $ with the T-duality direction $y_2'$  \johnt.
More generally, 
if we start  with the solution  $1_{NS} +1_R$ 
(parametrised by  the 
 $SL(2,R)$   angle $\theta$) 
  and  apply T-duality at an angle $\vp$ 
we get,  after lifting  the background to $D=11$,  the 2-brane solution where 
$y_{11}$  and  $y_1'$ are rotated by $\theta$.
The resulting  boosted and rotated 
2-brane solution is parametrised by the harmonic function 
 and the two  angles $(\theta, \vp)$.

It may seem that there is a  correspondence  between rotation in type IIB
theory and   boost  in $D=11$ theory:  a rotation and T-duality applied to $1_R$ 
 leads to the same type IIA configuration $2+0$  as  a 
finite boost and dimensional reduction applied to the  $D=11$ 2-brane. 
This relation is not, however,  universal: for example, 
T-duality applied to rotated 3-brane gives  the  $4+2$ solution  \mye\
which is the reduction of  static 
$5+2$ configuration \iz\ discussed below.
At the same time, 
the reduction of finitely boosted M5-brane $5\ma$  gives the 
 $5+0$ non-marginal configuration in $D=10$  \rust\
which is not T-dual to a rotated type IIB solution.
Another example is $6+4$ configuration 
 which is T-dual to rotated $5_R$ and
 is a dimensional reduction of  $7_{KK} + 5$, i.e. of a 1/2 supersymmetric
non-marginal configuration which is 
an interpolation  between  the   $D=11$ 
`Kaluza-Klein  7-brane' (or a KK monopole  \townelev)
 and a  5-brane.

Instead of boosting   M-branes  one can also rotate them. The reduction 
along the rotated direction 
also produces  non-marginal configurations in $D=10$  \refs{\rust,\papcost}.
For example, a plane wave along generic  cycle of 2-torus
in $D=11$ leads to finitely boosted 0-brane  $0\ma$ in  $D=10$
and 2-brane reduced along rotated direction becomes 
$2+1$ bound state in $D=10$. The T-duality relations in $D=10$ are 
$2+1 \to\  1_{NS} + 1_R \to\  0\ma$. In general, finitely 
boosting a RR $p$-brane  $p_R$ 
smeared in one transverse direction  and applying T-duality along this direction 
leads to  $(p+1)_R + 1_{NS}$ configuration, i.e.
a  non-marginal bound state of a RR  $(p+1)$-brane and a fundamental string \refs{\rust,\papcost}.

This illustrates how  some of  U-duality parameters in $D=10$  
are  simply the parameters of 
 coordinate transformations (boosts and rotations)
 in  $D=11$: 
 in general, a coordinate transformation
of a $D=11$ solution and  its dimensional reduction leads to the same type IIA 
configuration  as  certain  coordinate  transformation and T-duality 
applied to a type IIB solution.

Turning to more complicated examples, 
let us consider  the  non-marginal solution  $5+2$  \iz\ 
depending on  harmonic function
and one extra  angle $\t$ (in the single-center case  it is parametrised by  two 
charges).  It 
 `interpolates' 
between the  basic marginal $D=11$ solutions
-- the   2-brane ($\t=0$)   and  the 5-brane ($\t={\pi \ov 2}$).
The  corresponding metric is 
\eqn\izzi{
d s^2_{11} =  H^{1/3} \td H^{1/3}
 \big[  H\inv    (-  dt^2  + dy_1^2 + dy_2^2) 
 +  \td H\inv    (dy^2_3 + dy^2_4 + dy^2_5)  +  dx_m dx_m \big] \ ,   
 }
where $ \td H = 1 + {\rm sin}^2\t\  (H-1) $.

The  (static)    marginal configurations   with  two 
harmonic functions 
 are 
$2\bot 2$, $2\bot 5$ and $5 \bot 5$.  
To  find how to embed   $2\bot 2$ and 
 $2\bot 5$ into  a  single family of $D=11$  solutions
 let us   note that the reduction of $2\bot 5$ to $D=10$ is $2\bot 4$
 which is T-dual to $1 \bot 3$ or $2\bot 2$. Let us start  with
 $1 \bot 3$  with one  extra isometry ($y_5$) and  do  T-duality along the 
 rotated direction in the $(y_4,y_5)$ plane, where $y_4$ is one of the 3-brane's
 directions. If the rotation angle is zero, i.e. the  T-duality is 
 along $y_5$, we get
 $2\bot 4$ which lifts up  to $2\bot 5$. If the angle is ${\pi \ov 2}$ we get $(2\bot
 2)_1$  which is lifted to $2\bot 2$.
  The resulting $D=11$ background  which  interpolates between
 $2\bot 2$ and $2\bot 5$ thus has the following metric
 \eqn\eee{ ds^2_{11} = \td H_3^{1/3}H_3^{1/3} H_1^{1/3} 
[ - H_1^{-1}H_3^{-1} d t^2 +   H_1^{-1} dy_1^2 + H_3^{-1} (dy_2^2 + 
dy_3^2)  
}
$$  +\ \td H_3^{-1}H_3^{-1} d y_5^2
 +  \td H_3\inv (dy_4^2 +  dy^2_{11})+ dx_m dx_m ] \ , 
$$
where $\td  H_3 = 1 +{\rm sin}^2\vp (H_3-1) $  (in the single-center case 
$H_i = 1 + {Q_i/x^2}).$
This solution can be generalised to include 
 two more angles that will `connect'
$2\bot 2$ to $2+5$.  As a  result,  one finds 
   a family of 1/4 supersymmetric non-marginal
$D=11$ backgrounds 
which is parametrised by two independent harmonic functions and 
three  angles, and which contains the marginal configurations 
  $2+5$, $2\bot 2$  and $2\bot 5$ as special cases
(equivalent  $(2+5)\bot 2$ solution  appeared  in \costa).
Similar construction can be carried out by starting with
 $5\bot 5\bot 5$ configuration 
and finding a non-marginal family of solutions that will include
$2\bot 2\bot 2$,\ $2\bot 2\bot5$ and $5\bot 5\bot 2$ as special 
limiting cases.
Some 
examples of  composite  non-marginal $D=11$ configurations  with 1/4 and 1/8
of supersymmetry 
($(2+ 5)\bot (2+5)$, $(2+ 5)\bot (2+5)\bot (2+5)$, 
 etc.) were constructed in \costa.

To find a non-marginal 
1/4 supersymmetric $D=11$ 
 background which will include $2\bot 2$ and $5 \up$  as limiting cases
let us start with  the  $D=10$ type   IIA
$2\bot 2$  solution  with $(y_1,y_2)$ and $(z_1,z_2)$
as the internal spaces of the two 2-branes
and apply T-duality  twice in the  angled $(\vp,\psi)$ directions in the  planes
$(y_1,z_1)$ and $(y_2,z_2)$.\foot{The special case of this construction 
when the harmonic function of one of the two 2-branes was set equal to
one  was first considered in \mye. Another obvious generalisation is to 
apply T-duality at angles to $2\bot 2\bot 2$ configuration.
In the case when only one of the three harmonic functions is non-trivial
(i.e.  when one  2-brane is  `smeared' over 4 orthogonal directions) 
one finds the 1/2 supersymmetric non-marginal bound state
`$6+4+2+0$' considered from  D-brane point of view in 
 \refs{\lif,\gil}.
Related  non-marginal $D=11$ 
configurations were discussed in \ban.} 
Lifting the resulting 
 $D=10$ configuration (which can be denoted  symbolically as 
 $4 + 2 \bot 2+ 0$ or as $(4+2+0)\pa 0$)\foot{Such
 1/4 supersymmetric non-marginal 
bound state was considered in D-brane description in \lif\  where
  the existence 
of the corresponding supergravity solution was  also conjectured.
It was noted  there  that  taking a  4-brane with a self-dual 
 $F_{mn}$ 
background  (i.e.  with only one independent parameter) 
 and adding a 0-brane one finds a  trivial static potential, 
 implying that 
 there should  exist  a   1/4 supersymmetric  BPS  bound state of this generalised 
 `4-brane'  with an extra 0-brane.}
to $D=11$  gives  the following  1/4 supersymmetric 
background parametrised 
by  two harmonic functions and two rotation angles
\eqn\twoo{
ds^2_{11}= H_\vp^{1/3}H_\psi^{1/3} [ -H_1^{-1}H_2\inv d t^2
 +  H_\vp\inv (dy_1^2 + dz^2_1)
 +  H_\psi\inv (dy_2^2 + dz^2_2)
 }
 $$
+ \ H_1 H_2  H_\vp^{-1}H_\psi^{-1} (dy_{11} - A dt)^2  
       + dx_m dx_m ] \ , 
$$
$$
H_\vp \equiv  1 + (H_1-1) \coss\vp + (H_2-1) \sins\vp   \ , 
\ \ \
H_\psi \equiv  1 + (H_1-1) \coss\psi + (H_2-1) \sins\psi  \ ,  $$
$$
A= H_1\inv \sin\vp \sin\psi + H_2\inv \cos\vp\cos\psi \ , 
$$
$$
C_3=  (H_1-H_2) [ \sin\vp\cos\vp  H_\vp\inv dy_1\wedge 
 dz_1  -   
\sin\psi\cos\psi  H_\psi\inv dy_2\wedge  dz_2 ] \wedge dy_{11} 
+ ... $$
In the single-center  case $H_i = 1 + Q_i/x^3$,\  ($i=1,2,\vp,\psi$), 
and 
$\ Q_\vp= Q_1 \coss\vp + Q_2 \sins\vp, \ 
Q_\psi=Q_1 \coss\psi + Q_2 \sins\psi$.
We have written down explicitly only those  terms in $C_3$ 
which vanish  for $\vp=\psi=0$. 
Other terms  (which are  found using the 
T-duality transformation rules in the RR sector \berg) 
include  the $2\bot2$-type  structures
($H_1\inv dy_1\wedge 
 dz_1 \wedge dt  + H_2\inv dy_2\wedge 
 dz_2 \wedge dt$)
   as well as  a `magnetic'  5-brane type term.
 
Some special cases of the solution \twoo\ are:\ 
 $\vp,\psi=0:  \ 5 \up;  $ \ 
 $\vp=0,\ \psi={\pi \ov 2}:  \  (2\bot 2)_1;  $ \ 
 $H_2=1, \ \psi=0 :  \  2+ 5; $ \ 
$H_2=1, \ \psi={\pi \ov 2} :   \ (2\ma)_3 $.
Other special cases are found by using the 
 symmetries between $Q_1,Q_2,
 \vp,\psi$.
For example, the metric 
of $ (2\ma)_3$ (a 2-brane smeared in 3 isometric directions 
and finitely boosted along of them) is  indeed given by 
 \twoo\  with 
$\psi={\pi \ov 2}, $  $H_\psi=1, \ H_2=1$  
\eqn\opop{
ds^2_{11}= H_\vp^{1/3} [- H_1^{-1} d t^2  +  H_\vp\inv (dy_1^2 + dz^2_1)
 + dy_2^2 +dz^2_2 
} 
$$  + \  
H_1 H_\vp\inv (dy_{11} - H_1\inv \sin\vp\ dt)^2 + dx_m dx_m ] \ . $$
We thus  get an interpolation between the 1/2 supersymmetric 
$2+5$ configuration and  the   transversely boosted 2-brane in space with three 
extra  isometries $(2\ma)_3$, 
 and the 1/4 supersymmetric  infinitely 
boosted 5-brane $5
\up$ and  $2\bot 2$  in  the space with one  extra isometry.  
The supersymmetry is increased only  when  one of the two independent 
harmonic functions is set equal to 1.
The appearance of 5-brane is not surprising as $2\bot 2$ 
is T-dual (with T-duality applied twice along one of two 2-branes)
 to $4+0$ which  can be lifted up to $5 \up$ in $D=11$.

As  already mentioned above, it would be important to  describe  the rules
of constructing similar complicated non-marginal solutions directly in
terms of  $D=11$  theory. In view of the relation via dimensional reduction, 
the  set of  non-marginal configurations in $D=11$ is 
 in one-to-one correspondence
with the set of non-marginal configurations in $D=10$.
However,  in contrast to 
the  $D=10$  case,   the action of 
T-duality  on the $D=11$  set remains to be understood.

\newsec{More general `localised'  configurations of  branes in 
$D=10$ and $D=11$}
\subsec{String localised on 5-brane and related solutions }

It was noted 
in the previous sections that 
 essentially all marginal (and thus also non-marginal)  composite 
BPS configurations of branes in $D=10$ can be constructed by applying 
S- and T- dualities to the basic NS-NS  backgrounds  $1\up$, or $1\pa 5$,
or $1\pa 5\up$ which  correspond to exact conformal sigma-models.
These backgrounds are parametrised by several harmonic functions
satisfying the {\it flat-space}  Laplace equations. 
As we shall discuss below, starting with more general 
fundamental string type NS-NS backgrounds 
parametrised by  functions satisfying {\it curved-space} Laplace equations
one is able to construct more general composite $p$-brane configurations
by applying S-duality and 
 T-duality in  isometric directions.
While the  intersecting brane  solutions in  \refs{\papd,\tset} 
and the previous sections 
were isometric in all internal directions of the branes
(i.e. the position of the intersection was `smeared' over the branes)
and thus can be interpreted also as single anisotropic $p$-branes \guv\
these more general backgrounds 
correspond to `localised' intersections.

As was shown in \horts,  the following  string sigma-model 
 ($u,v=z\mp t$)
\eqn\sigmm{
L= H_1\inv (X) \del u\bar \del v  + L_\bot (X)  - \ha \a' 
 \sqrt{ g_2}R_2 \ln H_1 (X) \ , }
is  conformal   to all orders in $\a'$ provided the transverse theory
defined by $L_\bot  = (G_{ij} + B_{ij}) \del X^i \bar \del X^j 
+ \a' \sqrt{ g_2}R_2 \p_\bot(X) $ is  conformal 
and $H_1$ satisfies the  marginality condition, 
or the generalised Laplace equation, 
$ \nabla^i (e^{-2\p_\bot} \del_i) H_1  + ... =0.$
Dots stand for higher-order terms 
which  are absent 
when the transverse theory is 
$(4,4)$ supersymmetric 
as will always be the case in the examples discussed below.  
The `chiral null model'  \sigmm\ 
admits  a generalisation where
one includes also the terms like $ K(X) \del u \bar \del u $ and $A_i (X) \del u \bar \del X^i$.

The standard  fundamental string  solution \dgh\ corresponds to 
the trivial choice $L_\bot = \del X^i \bar \del X^i$ so that $H_1(X)$ 
is a flat-space  harmonic function.
Examples of more complicated  solitonic 5-brane-type \chs\ 
 choices of exact (super)conformal $L_\bot$
where considered in \refs{\CT,\TT}.
The resulting  solitonic  backgrounds
may be interpreted as a string lying on a 5-brane superposed with a Kaluza-Klein monopole \refs{\CT},  a string lying on a 
solitonic 5-brane \TT, or a string lying on a superposition of two 
solitonic 5-branes \TT. 
In the simplest case when the 8-dimensional  transverse space
part of \sigmm\  represents a  5-brane  wrapped over a flat 4-torus
one finds the following  background ($X=(y_n,x_m)$, \ $n,m=1,...,4$, 
 cf.\fio)  \TT\ 
\eqn\qeq{
ds^2_{10} = H_1\inv (x,y) (-dt^2 + dz^2) +  dy_n dy_n  + H_5(x) dx_m dx_m \ , }
$$
dB= dH\inv_1 \wedge dt\wedge dz  + *dH_5  \ , \ \ \ 
e^{2\phi} = H_1\inv   H_5   \ , $$
where $(z,y_n)$ are the internal dimensions of the 5-brane,  
 $H_5(x)$ is the harmonic function ($\del^m \del_m H_5 \equiv \del^2_x H_5=0$)
defining the position of 5-brane(s) and the string function $H_1(x,y)$ 
 satisfies\foot{I am grateful to J. Maldacena for pointing out 
an error  in this equation in the original version 
of \TT\ and  useful discussions of related  localised $p$-brane 
solutions.}
\eqn\uuu{
 [  \del^2_x +  H_5(x)  \del^2_y] H_1 (x,y) =0 \  . }
The same equation  should be satisfied by the 
function $K(x,y)$ of the  longitudinal wave.
An obvious special solution  is found by taking $H_1$  as 
 a product (or a  sum) of  the two  special 
 harmonic functions,  
$H_1 (x,y) = H_1(x) H_1'(y)$. 
Such factorised solution does not match, however, onto a (sum of) 
 delta-function  string source(s) $-\mu \delta^{(4)} (x) \delta^{(4)} (y)$
which should be present in  the r.h.s. of \uuu\ 
 for a localised string  
solution. Though it is not important for 
for what  follows, it is natural to assume
  that the solution  $H_1(x,y)$
of \uuu\  should be  chosen  in such a way 
  that in the limit $H_5\to 1$ it becomes a free fundamental 
string one with the 
harmonic function having  isolated singularities in the 
8-dimensional transverse space. Unfortunately, it turns out that 
such a solution
does not have a simple expression 
in   terms of elementary functions
 even for the one-center choice of the 
5-brane function, $H_5= 1 + Q/x^2$. 

Ref.  \TT\  
concentrated 
on the  special solution    for which  $H_1$ does not depend on the 
  5-brane coordinates $y_n$ 
 transverse to the string. The main reason was that 
such a  background directly corresponds  to  an extremal 
black hole in $D=5$ upon  dimensional reduction.
In this special case ($H_1(x,y) = H_1(x),\  \del_x^2 H_1=0$)  
the string is smeared over the 5-brane
so that the background has 5  spatial 
isometries and
is parametrised by the  flat-space harmonic functions $H_5(x)$ and $ H_1(x)$.
More general solutions  representing a  string localised
on the 5-brane, i.e. with $H_1$ having   non-trivial 
dependence on $y_n$ were  recently discussed 
 in \refs{\HM,\US}.
Similar `localised' generalisations  exist for the conformal 
models in \refs{\CT}  which,  in the `smeared' case,  describe
 extremal black holes in $D=4$ which have regular horizons.

The localised solutions have the same amount of supersymmetry 
 and the same BPS 
marginality property as the
`smeared' ones. These properties 
are universally determined by the special holonomy 
of the generalised  connection of the corresponding 
chiral null model \sigmm\ (implying also its $(4,4)$ supersymmetry
 in the case of type II superstring  theory).
They are  the consequences of
 the special choice  of the transverse 
theory (and of the chiral null structure of \sigmm) 
and do not depend on the form of  $H_1(x,y)$.
 However, in contrast to the delocalised 
solution where the two harmonic functions  $H_1(x)$ and $H_5(x)$
appear on an equal footing and  specify  the position(s)
of the  string(s) and 5-brane(s) (which,  in general,  are independent 
and arbitrary),  the roles of $H_5(x)$ and $H_1(x,y)$ are
 obviously
 asymmetric  in the localised  case \uuu.
To distinguish 
 the localised intersections from  the smeared ones 
we shall put `hats' on the 
  symbols $\pa, \bot$,  i.e. 
the configuration described by  \qeq\
will be denoted as $1_{NS} \pab 5_{NS}$.

Applying S-duality to $1_{NS} \pab 5_{NS}$
  as a type IIB solution 
  we find the $1_{R} \pab 5_{R}$   configuration  
 describing 
a RR string localised on RR 5-brane (which generalises the 
`smeared' configuration used in  \CM).
T-duality along the isometric $z$-direction gives the  $4\pab 0$
configuration with 
 the position of the 0-brane (determined by $H_1(x,y)$)
 being localised on the 4-brane. 
If we assume that $H_1(x,y)$ does not depend on one of the four 5-brane coordinates,  i.e. that the string is smeared over,  e.g.,  $y_1$, 
then applying T-duality along $y_1$ we find the $2\botb 4$ 
type IIA configuration in which the position of the intersection string is 
localised on the 4-brane ($z,y_2,y_3,y_4$)
 but {\it not}  on the  2-brane  ($z,y_1$).
 This asymmetry becomes even more apparent if we 
delocalise the solution also in $y_2$ and again apply T-duality 
along the resulting isometric direction. 
Using T-duality transformation rules \berg\ one  finds the 
following  $3\botb3$
configuration (which generalises the `smeared'  $3\bot 3$ solution 
   \TT) 
 \eqn\fwww{
ds^2_{10}=
(H_3 H'_3 )^{1/2}
\big[ (H_3 {H}'_3)^{-1}  (- dt^2  + dz^2)  
+  {H}'^{-1}_3  (dy_1^2 + dy^2_2) } $$ 
  +\  H\inv _3  (dy^2_3 +dy_4^2)  +  dx_m dx_m \big] \ ,   $$
$$ dC_4 
= dt \wedge dz (dH\inv_{3}  \wedge dy_1 \wedge dy_2 
+   {H}'^{-1}_3 \wedge dy_3\wedge dy_4)  $$ $$
 +\   *d_x H_{3}\wedge dy_1 \wedge dy_2 
+ *dH'_{3} \wedge dy_3 \wedge dy_4
+  H_3' *d_y H_3 \wedge dx_1 \wedge dx_2 \wedge dx_3 \wedge dx_4  \ , $$
$$ H_3 = H_1 (x,y_3,y_4), \ \ \ \  \  H'_3=H_5 (x)  \ , \ \ \ \ 
 \ e^{2\p} =1 \ , $$
where $*d_y  \equiv dy_4  \del_{y_3}  -dy_3  \del_{y_4}  $.   
It seems unlikely that there exists a
generalisation of this solution  
 in which 
the two  3-brane functions $H_3$ and $H_3'$ appear symmetrically 
(i.e. $H_3'$ depends on $y_1,y_2$)
and which still has a  BPS property. 
T-duality along $z$-direction then would give 
the  localised $2\botb 2$ solution smeared  only in one transverse direction.\foot{Such   solution would not 
have an obvious  analogue in the NS-NS sector.
The structure of the metric of the intersecting  $p$-brane solutions 
seems to be rather  rigidly fixed by the BPS condition \tser.
Also, eq. \uuu\ does not seem to admit a non-trivial generalisation
to a system of two equations for $H_1$ and $H_5$, 
both  depending  on $y_n$ (related observations were  made
 in \gaukas).}

More localised intersections can be constructed by first 
smearing in some of the transverse directions and then using
 T-duality.
For example, applying T-duality to  the above $3\botb3$ 
configuration along $x_1$ one finds $4\botb 4$ 
solution where the intersection 2-brane
 is localised only on one of the two 
4-branes. Another  example is  obtained  by starting with $1\pab 5$
smeared in one transverse direction, i.e. 
having  isometry in $x_4$ coordinate in \qeq.
T-duality along $x_4$ then 
 converts  the solitonic 5-brane part of \qeq\ 
into the `Kaluza-Klein 5-brane' (or KK monopole) part, which is  a 
purely gravitational (Euclidean Taub-NUT) background  
\eqn\qeqi{
ds^2_{10} = H_1\inv  (-dt^2 + dz^2) +  dy_n dy_n 
 + H_5\inv  (d x + \B_i dx_m)^2  +  H_5  dx_m dx_m\ , }
$$
dB= dH\inv_1 \wedge dt\wedge dz   \ , \ \ \ 
e^{2\phi} = H_1\inv   \ , \ \ \ \    d\B= *dH_5 \ ,  \ \ \  H_1=H_1(\vx,y)\ , \ \  H_5 =H_5(\vx) .  $$
We  have set $x_n=(x_m,x_4)$ and used 
 $x$  to denote  the  coordinate dual to $x_4$.
Starting instead with $5_R\pab 1_R$ solution and applying T-duality along $x_4$  and $y_4$ we find  the $5_R\botb 3$ (and, 
 by S-duality,  $ 5_{NS}\botb 3$) configuration, i.e. the  
intersection of a   5-brane  (which is smeared over  $x$) 
with a 3-brane over a 2-space 
 which is localised on the 5-brane.\foot{One  may  
 consider a periodic array of 5-branes in $x$-direction, 
all intersected by a 3-brane having  $x$ as  the  dimension transverse
to the 5-brane. 
  Similar configurations (but  with  localisation   in $x$)
were discussed in \hanw.}
  More general 
 backgrounds  including these as special cases 
will be considered   below.

Lifting  the $1\pab 5$ solution \qeq\
to $D=11$  by adding the  isometric direction $y_{11}$
one finds the  M-brane intersection $2\botb 5$
where the intersection string is localised on the 5-brane $(z, y_1,y_2,y_3,y_4)$
but not  on the 2-brane $(z, y_{11})$,  
\eqn\qequ{
ds^2_{11} = H_2^{1/3}  H_5^{2/3} [H_2\inv 
  H_5\inv 
 (-dt^2 + dz^2)    +   H_2\inv dy_{11}^2 + 
 H_5\inv   dy_n dy_n  +  dx_m dx_m ]\ ,  }
$$ dC_3 
 = (dH_2\inv  \wedge dt \wedge dz  +  *dH_5  ) \wedge dy_{11}                        \ , \ \ \  \ \  H_2\equiv  H_1(x,y)\ , \ \ H_5=H_5(x) \ .  $$
Similarly, lifting  the configuration
$4\botb 4$ (T-dual to $3\botb 3$ \fwww\ as 
mentioned  above)
  to $D=11$
one finds the $5\botb 5$ solution, again with asymmetric localisation of the intersection 3-brane on only one of the two 5-branes.\foot{This 
localised solution may be of interest  in  the context of 
 discussions \refs{\hankl} of
3-branes as intersections of two M5-branes.}
Starting with $3\botb3$ smeared in $y_3$ and applying T-duality along this coordinate
one finds the $2\botb  4$ solution with 
the intersection string localised on the 2-brane
($H_4=H_3(x,y_4)$, $H_2=H_2(x)$). Lifting this solution to $D=11$
gives the $5\botb 2$ solution similar to \qequ\
where $H_2=H_2(x), \ H_5=H_4(x,y_4)$, i.e. the intersection is localised on 
the  
2-brane instead of the 5-brane.\foot{Let us note also that applying T-duality to $3\botb3$ along the intersection 
string direction gives $2\botb 2$ solution with one transverse isometry, which is lifted to 
a $2\botb2$ solution in $D=11$ with two transverse isometries.
  There does not seem to exist a   similar localised $2\botb 2$ solution with no transverse isometries.}
These  1/4  supersymmetric localised   
intersecting M-brane solutions were 
independently found  by  J. Gauntlett  \gaukas.

Localised  
BPS configurations of branes parametrised by three 
 harmonic functions can be constructed in an  
analogous way by starting with a
generalisation  of \qeq\ with an additional 
 function $K =H'(x,y)-1$ satisfying 
\uuu\  and representing the longitudinal momentum 
wave along the string localised on the 5-brane, $5\pab 1\up$. 
 One can then construct various intersecting 
configurations by relaxing localisation 
in some of the internal 5-brane 
coordinates and/or smearing in some of the transverse  dimensions 
 and  applying 
S- and T-duality. Lifting the resulting backgrounds 
 to $D=11$  leads 
to 1/8 supersymmetric  intersections of M-branes 
with (varied amount of) localisation, e.g., $2\botb 5\botb5$, etc.

\subsec{String localised on intersection of two 5-branes and related solutions }

One  straightforward  generalisation of the above discussion 
is obtained by  replacing 
 the  product of the 4-torus and curved `5-brane' 4-space  which was used 
as  an  8-dimensional transverse 
conformal theory in \sigmm,\qeq\
by the direct product of the two 5-brane  theories \khu\ 
(which is  obviously conformally invariant
being described by the sum of  two independent 
conformal sigma-model actions).\foot{It is possible also to 
consider more general non-direct-product 8-dimensional 
 conformal models based on hyper-K\"ahler metrics (and their 
generalisations  including antisymmetric tensor background). 
This  leads to more general 
intersecting brane solutions   constructed in \gaugi.}
The resulting  conformal 
background   is  \TT\
\eqn\qeqty{
ds^2_{10} = H_1\inv (x,y) (-dt^2 + dz^2) + H'_5(y)
 dy_n dy_n  + H_5(x) dx_m dx_m \ , }
$$
dB= dH\inv_1 \wedge dt\wedge dz  + *dH_5 +  *dH'_5 \ , \ \ \  \ \ 
e^{2\phi} = H_1\inv   H_5  H'_5   \ , $$
where  the harmonic functions 
$H_5(x)$ and $H_5'(y)$  ($ \del^2_x H_5 =0, \ \del^2_y H_5'=0$)
define the positions of the two 5-branes $(z,y_n)$ and $(z,x_m)$
 and the string function  $H_1(x,y)$ satisfies
\eqn\uuu{
 [  H_5'(y) \del^2_x +  H_5(x)  \del^2_y] H_1 (x,y) =0 \  . }
In the special case of  $H_1=1$ the background \qeqty\ 
 represents the configuration of 
 two 5-branes   which share the string direction 
and are `localised' with respect to each other. 
 We shall denote this 
configuration  as  $5_{NS} \cap 5_{NS}$ following \gauk\
where this `overlapping 5-brane' interpretation 
of the solution of \khu\ was suggested  and it 
was  lifted to  $5\cap 5$ in $D=11$.  
Another special case  with $H_1= H_1(x) H_1'(y)$  
corresponds to the  `dyonic  string' 
generalisation  of the solution of  \khu\ 
 which   was found  in \dufe. 

We shall use the notation 
$5\cap 5\pab 1$  for the  general NS-NS solution \qeqty,\uuu.
Starting with this configuration and applying S-duality  and T-duality
along $z$ we get the related solutions $5_R\cap 5_R\pab 1_R$ 
and $4\cap 4\pab 0$. Lifting  $5\cap 5\pab 1$ to $D=11$ 
we find the $5\cap 5\botb 2$  M-brane configuration  
which generalises the $5\botb 2$ solution  \qequ\ to the presence of an
additional 
5-brane (and generalises $5\cap 5$ of \gauk\ to the presence of the 
2-brane)
$$ds^2_{11} = H_2^{1/3} H_5^{2/3}  H_5'^{2/3}[(H_2 H_5 H'_5)\inv
(-dt^2 + dz^2)  +   H_2\inv  dy_{11}^2$$  
\eqn\qqu{
   +\  H_5\inv   dy_n dy_n  +  H'^{-1}_5  dx_m dx_m ]\ ,  }
$$ dC_3   
 = ( dH_2\inv  \wedge dt \wedge dz  +  *dH_5  + *dH'_5  ) \wedge dy_{11}                        \ , $$
$$H_2=H_1(x,y)\ , \ \ \ \  H_5=H_5(x) \ , \ \ H'_5=H'_5(y) \ . $$
Other related  $D=10$ solutions 
can be constructed by relaxing localisation in some of  the $(x_n,y_m)$ coordinates and applying T-duality along these isometric directions.
For example, 
let us split the coordinates  in the 3+1 way, 
$x_m=(x_i, x_4), \ y_n=(y_i, y_4), \ i=1,2,3$, 
 and assume that the functions $H_1,H_5,H_5'$
do not depend on $x_4,y_4$.
Starting with $5_R\cap 5_R\pab 1_R$  type IIB solution
delocalised in $(x_4,y_4)$ 
$$ ds^2_{10}  = (H_1 H_5 H_5')^{1/2}[(H_1 H_5 H_5')^{-1}
   (-dt^2 + dz^2)
 + H\inv_5 (dy^2_4 +  d\vy^2)+   {H'^{-1}_5}(dx^2_4 + d\vx^2)  ]  \ ,  
  $$ 
\eqn\erw{
e^{2\phi} = H_1 ( H_5 H_5')^{-1}\ , \ \  \ \  \  
dC_2=  dH\inv_1 \wedge dt\wedge dz  + *dH_5\wedge dx_4
 +  *dH'_5\wedge dy_4      \ ,  
    }
$$ H_1=H_1(\vx,\vy) \ , \ \ H_5=H_5(\vx) \  , \ \ H_5'=H_5'(\vy) \ , $$
and applying T-duality along $x_4$ and $y_4$ one finds the 
 $5_R\cap 5_R\botb  3$ configuration  with the metric 
which has the expected `harmonic function rule' form 
$$ ds^2_{10}  = (H_3 H_5  H_5')^{1/2}[(H_3 H_5 H_5')^{-1}
   (-dt^2 + dz^2) $$
\eqn\wyv{
+  \  (H_3 H_5)\inv dx^2  +  (H_3 H'_5)\inv 
dy^2   +   H\inv_5   d\vy^2   + {H'^{-1}_5} d\vx^2
 ]  \ , }
$$H_3=H_1(\vx,\vy) \ ,\ \ \  H_5=H_5(\vx) \  , \ \ \  H_5'=H_5'(\vy) \ . $$
Here $x,y$ are dual to $x_4,y_4$ and the  
 3-brane coordinates  are $(z,x,y)$.
 Each of the two 
5-branes $(z,x,y_i)$ and $(z,y,x_i)$ (which share  one string direction)
 intersects with the 3-brane over a 2-space (note that $x,y$ effectively
interchanged places compared to $x_4,y_4$).
The 3-brane is localised on each of the 5-branes only in 2 out of 3 coordinates.
 This generalises the  $5_R\botb 3$ solution mentioned  above.

 S-duality  then leads to  the  solution
$5_{NS}\cap 5_{NS}\botb 3$ where the RR 5-branes are replaced by the 
NS-NS ones.
It is possible also to construct a solution
representing the  S-`self-dual' configuration  
$5_{NS}\cap 5_{R}\botb 3$
with the two different types of 5-branes, 
which intersect  each other and  the 3-brane 
over a 2-space 
(the existence of such  BPS  
configuration was pointed out  in \hanw).
The corresponding  background can be constructed by applying U-duality to 
$5_{NS}\cap 5_{NS} \pab 1_{NS}$
delocalised in $x_4$ and $y_4$. 
Indeed, T-duality along $x_4$  transforms  first  
 $5_{NS}$  ($z,\vy,y_4)$ into KK 5-brane  $5_{KK}$ (described by 
euclidean Taub-NUT metric, cf.\qeqi).  S-duality then 
converts $5_{KK}\cap 5_{NS} \pab 1_{NS}$
into $5_{KK}\cap 5_{R} \pab 1_{R}$
 (being a purely gravitational background, 
$5_{KK}$ is invariant under S-duality).
Applying T-duality twice along $x_4$ and $y_4$ 
leads to $5_{NS}\cap 5_{R}\botb 3$ solution.

Alternatively, one may start  with its $D=11$ counterpart 
$5\cap 5 \botb 2$ \qqu\ 
`smeared' in $x_4$ and $y_4$ directions 
so that it has 
 $dC_3= (dH\inv_1 \wedge dt\wedge dz  + *dH_5\wedge dx_4
 +  *dH'_5\wedge dy_4)\wedge dy_{11}$.  
Since $x_4$ is now an isometry,  one may reduce this  solution 
down to $D=10$ along $x_4$ 
obtaining the $4\cap 5 \botb 2$ type IIA solution. 
Applying T-duality along $y_4$ then leads to  the 
$5_{NS}\cap 5_{R}\botb 3$ type IIB  background
$$ ds^2_{10}  = (H_3 H_5')^{1/2}  H_5 
[(H_3 H_5 H_5')^{-1}
   (-dt^2 + dz^2 + dy^2)  $$ 
 \eqn\wyve{   + \     H\inv_3 dx^2   + H\inv_5   d\vy^2 
+    {H_5'^{-1}} d\vx^2
]  \ ,   } 
$$ 
 e^{2\phi} = H_5 H_5'^{-1}, \ \ \ \
 dB = *dH_5\wedge dx   \ , \ \ \ \
   dC_2=   *dH'_5\wedge dx\ ,  $$  $$ 
dC_4=  dH^{-1}_3 \wedge dt\wedge dz \wedge dx\wedge dy 
+  H_5 * d_{\vy} H_3 \wedge dx_1 \wedge dx_2\wedge dx_3
+  H'_5 * d_{\vx} H_3 \wedge dy_1 \wedge dy_2\wedge dy_3
 , $$ 
$$  H_3= H_1(\vx,\vy) \ , \ \ \  \  
 H_5=H_5(\vx) \  , \ \ \  \ H_5'=H_5'(\vy) \ . $$
Here $y$ denotes the  coordinate 
dual to $y_4$, \  $x\equiv y_{11}$ and 
$* d_{\vx} =\ha  \ep_{ijk}  dx_j \wedge dx_k \del_{x_i} , \ $
$* d_{\vy} =\ha  \ep_{ijk}    dy_j \wedge dy_k \del_{y_i}.$
The   coordinates of the branes are $3$:  $(z,y,x)$, \ $5_{NS}$: $(z,y,y_i)$, 
\  $5_{R}$: $(z,y,x_i)$.
 S-duality maps this background into  itself with $ H_5 \leftrightarrow H_5'$,
$\vx \leftrightarrow \vy$. 
The  3-brane is localised only
 relative to the $(x_i,y_i)$  3-spaces of the two 5-branes, and
all branes are delocalised in the common transverse direction 
 $x\equiv y_{11}$.
 It is not clear if there exists a  similar 
static solution
describing   the branes  localised in $x$, i.e.  
 the  configuration considered in  \hanw.\foot{On possibility 
(suggested by the expressions for the moduli metrics in \hanw) 
is that 
 a solution localised in $x$  will have $H_5$ ($H_5'$) replaced by the sum
of harmonic function in $x$ and a harmonic function in $x_i$ ($y_i$), i.e. 
$H_5 = q |x-x_0| + {Q \ov | \vx - \vx_0|}$.
Such function is obviously a solution of 5-brane conformal invariance condition
in \qeqty\ 
($\del^m\del_m H_5=0$, with $x=x_4$) but it is not clear that such an ansatz is fully consistent as one is no longer able to apply T-duality in $x$-direction
to relate various configurations as was done above.}

For comparison, let us note 
that there exists  another marginal $ 5_{NS} \bot 5_R\bot 3$ 
type IIB configuration  which is  covariant under S-duality.
This is the delocalised intersection 
 where the 5-branes intersect over a 4-brane  and each intersects
the 3-brane over a 2-space.
This  configuration is 
T-dual (along  direction parallel 
to   $ 5_{NS}$) to $5_{NS}\bot 4\bot 4$  which is 
a dimensional reduction
of  the configuration $5\bot 5\bot 5$ of 
three orthogonal 5-branes in $D=11$ \refs{\papd,\tset}.
The metric  of  $ 5_{NS} \bot 5_R\bot 3$ is (cf. \wyve) 
$$ ds^2_{10}  = (H_3 H_5')^{1/2}  H_5 
[(H_3 H_5 H_5')^{-1}
   (-dt^2 + dz^2)  +  (H_5 H_5')^{-1}(dy_1^2 + dy^2_2 + dy^2_3) $$
 \eqn\wve{   + \     (H_3H_5)\inv  dy^2_4 
   + (H_3H'_5)\inv  dy^2_5  + dx_i dx_i ] \ , }
where $H_5, H_5', H_3$ depend only  on $x_i$.

\newsec{Actions  for  brane probes in backgrounds of 
composite $p$-brane configurations}

An  advantage of knowing
explicitly  the supergravity  backgrounds 
representing composite configurations of different type of branes
 is that one can easily determine the structure of classical  
actions of $p$-brane probes moving in closed string 
 backgrounds produced by  the corresponding  systems  of   brane  sources.
In the case of supersymmetric
D-brane configurations this determines the form of the second-derivative 
terms in the action which appear as 
 1-loop  corrections in the  open string theory description 
\refs{\polch,\bakas,\doug}. This 
classical approach is particularly useful in the
 case  when  (some of) the sources are the   NS-NS branes 
 for which there is no simple  analogue of a  perturbative 
D-brane description.

For example, 
 let  us  consider a fundamental   string probe 
moving in the $5_{NS} \cap 5_{NS} \pab 1$ background \qeqty.
Orienting the probe  along $(t,z)$  
and choosing the static gauge  ($X^a=\s^a$) 
one finds that the string action 
$$I_1=T_1 \int d^2\s[  \sqrt {- \det { (G_{MN}(X) \del_a X^M \del_b X^N)} }
+  \ha B_{MN}(X) \ep^{ab}  \del_a X^M \del_b X^N] $$
takes the following form
$$ I_1=T_1 \int d^2\sigma
  [ 
\big( - \det [H\inv (x,y) \eta_{ab}  +  H_5'(y) \del_a y_n \del_b y_n $$
$$ + \  H_5(x)
 \del_a x_m \del_b x_m  ]
 \big)^{1/2}   - H\inv (x,y) ] $$
 \eqn\ati{
 = \ha T_1 \int d^2\sigma[  \ H_5'(y) \del_a y_n \del_a y_n
 +  H_5(x)
 \del_a x_n \del_a x_n + ... ] \ . }
The  vanishing of the static potential indicates  that 
this is a  BPS configuration.
The  dependence on $H_1(x,y)$ cancels out 
in the second-derivative approximation. This is related to the 
absence of velocity-squared corrections to the force (i.e. to the 
flatness 
of the  moduli space)  in the system of parallel strings.
In general, the moduli space metric in \ati\
is the same as the 
8-dimensional hyper-K\"ahler metric in \qeqty\
(in the  single-center case $H_5=1 + Q/x^2, \ H_5'=1 + Q'/y^2$).
The same result is found by considering the 
S-dual situation -- a RR string probe moving  in $5_{R} \cap 5_{R} \pab 1_R$
background.

Let us now consider the bosonic terms in the action for a 3-brane probe 
moving in a type IIB supergravity background (see, e.g.,  
\pol)
\eqn\threea{
 I_3=T_3 \int d^4\sigma \big[
   e^{-\phi} 
    \sqrt {-\det (\hat  G  +  \F) }
    +{\textstyle  {1\ov 4!} } \ep^{abcd} \hat C_{abcd} 
+   \ha  \F^{*ab}  \hat C_{ab} 
+  \four  C \F^{*ab}  \F_{ab}  \big] 
 \ , }
where 
$ \F_{ab}= F _{ab}+ \hat B_{ab} ,  \  F=dA  , \
 F^{*ab} = \ha \ep^{abcd}F_{cd},$ \ 
$    \hat G_{ab} = G_{MN} \del_a X^M \del_b X^N,$ etc., 
and   $a=0,1,2,3$.\foot{To make this action manifestly covariant under $SL(2,Z)$ duality
\refs{\me,\greee}  one should add a $B\wedge C_2$ term so that 
$\F^{*ab}  \hat C_{ab}$  becomes $F^{*ab}  \hat C_{ab}$.
Equivalently, this corresponds to choosing   $\hat C_4$ as 
$\hat C_{abcd} = \hat  C'_{abcd} - 6 \hat B_{[ab} \hat C_{cd]}$
where $C'_4$ is invariant under $SL(2,Z)$.
This  subtlety will not be important in what follows 
as $ B\wedge C_2$ will vanish for the backgrounds we shall consider.}

It is important to note that  the gauge field  $A_a$ will 
not, in general, decouple from the background
and should  be taken into consideration
in discussing the low-energy (moduli-space) approximation.

For example, if the background is produced by an NS-NS 5-brane 
smeared over  one transverse ($x_4$) direction,  
the action  for a 3-brane probe positioned parallel to $(z,x,y$)
directions is (here $y_n=(y_i, y_4\equiv y), \ x_i =(x_i, x_4\equiv x)$) 
\eqn\iyi{ I_3=T_3 \int d^4\sigma 
   \sqrt {\det [\delta^a_{b} + \kappa^{ac}
   (\del_c y_i \del_b y_i + H_5\del_c x_i \del_b x_i 
     +  \F_{cb}) ] } \ , }
$$ \kappa_{ac} = diag(-1,1,1,H_5) \ ,  \ \   \ \  \F = F
 +  \B_i(\vx)  dx_i \wedge dx  \ , \ \ \  \ d\B= *dH_5  \ , \ \ \ H_5=H_5(\vx) \ .  $$
Expanding in powers of derivatives we get   
\eqn\utu{
 I_3=T_3 \int d^4\sigma 
   [1 + \ha \kappa^{ac}
   (\del_a y_i \del_c y_i + H_5\del_a x_i \del_c x_i)
     + \ha  \kappa^{ac} \kappa^{bd }\F_{ab}\F_{cd} + ... ] \ . 
}
Let us split the world-volume indices as $a=(k,3)$, 
 $k=0,1,2$,  take  $A_a= (A_k, A_3\equiv \t)$
and assume
 that all the  fields do not depend 
on $\s_3$ (equal to $x$ in the  static gauge). 
Then  $  \kappa^{ac} \kappa^{bd }\F_{ab}\F_{cd}   =  F_{kl}^2 
 + 2 H_5^{-1} (\del_k \t + \B_i \del_k x_i)^2 $ so that 
\eqn\tte{
I_3=  \ha T_3 \int d^4\sigma 
   \big(2+  \del_k y_i \del_k y_i   + H_5(\vx)\del_k x_i \del_k x_i
    +  H_5^{-1}(\vx)  [\del_k \theta  + \B_i (\vx)  \del_k x_i]^2 \big) + ...
\ ,      }
     where dots stand also for  the decoupled    $F_{kl}$-terms. 
 Keeping the component 
 $A_3= \t$ of the world-volume gauge field 
     is important  as it does not decouple from the background. 
This  leads  to the following  moduli space metric
(for  future comparison,  we  include also  the contribution
of the   decoupled  component $A_2\equiv \t'$)
\eqn\ouo{
 ds^2 =  dy_i^2 + d\t'^2 + H_5(\vx)  dx_i dx_i + H_5^{-1}(\vx) 
[d \t + \B_i (\vx)  d x_i]^2  \ , } 
$$ d\B= *dH_5  \ ,\ \  \ \ \ \ \
 H_5 = 1 +  \sum_s  {Q_s \over |\vx_s- \vx_{0s}|}\ . $$
Its  curved part is the same 
hyper-K\"ahler metric as in  the Kaluza-Klein 5-brane 
background (cf.\qeqi)
which is related  to the solitonic 5-brane  by T-duality 
along $x=x_4$. Here the role of the coordinate dual to $x_4$ is played
by the component $A_3=\t$ of the gauge field
($\t'$ corresponds to the dual of  $y_4$).
 This  should not be surprising 
 as T-duality applied to the  whole system 
including the probe  should transform   the corresponding 
gauge field component
into a   D-brane collective coordinate.  Related
 discussion 
appeared in \hanw\ where the 5-brane (or a collection of parallel 
5-branes)  was
  assumed to be localised in $x=x_4$, while here we are considering
 a `smeared'  case.

The same action 
is found if the $5_{NS}$ background is replaced by the $5_R$ one.
This follows from the $SL(2,R)$ covariance 
of the  3-brane action in a type IIB supergravity background \refs{\me,\greee}.
Here the  role of an `extra'   coordinate 
is played by the magnetic counterpart
of the `electric' gauge field variable $\t$.
Indeed, 
  in the RR 5-brane  background $B=0$ and so 
$\F=F$ but instead  there is   the  $F \wedge  \hat C_2$ coupling
with  $C_2= \B \wedge dx$. 
Adding  this term to the $H_5(\vx) \kappa^{ac} \kappa^{bd } F_{ab}F_{cd} $ one 
 coming  from  expansion  of the Born-Infeld action
(which has an extra factor  of $H_5$ compared to \tte)
and introducing $  \epsilon_{kls} \del_s \tt$ as the dual 
 `monopole' part of $F_{kl}$ (or performing $d=3$ duality, see below) 
one finishes, after decoupling  of  other
 gauge field components, 
 with  an  equivalent action, where  $\tt$ is playing  the role of $\t$, 
 so that   the moduli space metric contains 
  the  term 
 $H_5\inv (d\tt + \B_i dx_i)^2$.

The cases of more complicated  configurations like $5_{NS} \cap 5_{NS}$
or $5_{NS} \cap 5_R \botb 3$ are treated in a similar way.  
For example, putting  a 3-brane probe in the $5_{NS} \cap 5_{NS}\botb 3$
  background  
leads to the following action 
  (the 3-brane probe is oriented 
along $(z,x,y)$, cf. \wyv)
 $$ I_3=T_3 \int d^4\sigma\
   \big[ H^{-1}_3 \big( 
    \det [\delta^a_{b}  
    + \kappa^{ac} (   H_3 H_5' \del_a y_i \del_c y_i $$
 \eqn\oto{  +\     H_3 H_5
 \del_a x_i \del_c x_i  
  +   H^{1/2}_3\F_{ac})] \big)^{1/2}  
 -  H^{-1}_3  \big]  \ ,    }
where $H_3=H_1(\vx,\vy) ,  \ H_5=H_5(\vx) ,  \ H_5'=H_5'(\vy) $, 
$$ \ 
\kappa_{ab} =diag( -1,1,H_5,H_5')\ , \ \ \ \   \ \
  \F=F + 
\B \wedge dx +  \B' \wedge dy \     
 , $$
 and the $-H_3\inv $ term in the potential came  from the 
 $C_4$-background produced by  the  3-brane source.
As in \ati\ the dependence on the 3-brane source function  $H_3(\vx,\vy)$ 
disappears in the moduli space approximation.
 Introducing   $A_2 =\t'$ and $A_3=\t$
and decoupling the remaining components of the 
gauge field 
 one finds  the following moduli space metric 
\eqn\yuy{
  ds^2 =  H_5'(\vy) dy_i dy_i  +     H_5'^{-1}(\vy) 
[d \t' + \B'_{i} (\vy)  d y_i]^2 } $$ 
+\ H_5(\vx)  dx_i dx_i + H_5^{-1}(\vx) 
[d \t + \B_{i} (\vx)  d x_i]^2    \  ,   $$
which is the direct product of the two 4-d Euclidean Taub-NUT 
 metrics  in \ouo. This  8-dimensional hyper-K\"ahler 
 metric is related to the transverse metric of the 
  $5_{NS} \cap 5_{NS}$  background 
by T-duality in $x_4$ and $y_4$
(cf. \qeqi).

 Finally, let us consider the  most interesting
case of the 
`mixed'    $5_{NS} \cap 5_R \botb 3$ 
background \wyve\  which is `self-dual' under the $SL(2,Z)$. 
 The corresponding 3-brane action is (cf. \oto)
 $$
 I_3=T_3 \int d^4\sigma \big[  H^{-1}_3 \big(
    \det [\delta^a_{b}  
    + \kappa^{ac} ( H_3 H_5' \del_b y_i \del_c y_i+ H_3 H_5
 \del_b x_i \del_c x_i
  $$
\eqn\ioi{  + \     (H_3 H'_5)^{1/2} \F_{cb})] \big)^{1/2} 
-  H_3\inv  +  \ha 
     \F^{*ab}\hat C_{ab} + ...  \big] \ ,  }
where $H_3=H_1(\vx,\vy)$,   $(H_{5})_{NS}=H_5 (\vx)$,
 $ (H_{5})_{R}=H'_5 (\vy)$,  
$$\kappa_{ab} =diag( -1,1,1, H_5 H_5')\ , \ \ \ \
 B = \B_i dx_i \wedge dx\  ,
 \ \  \  C_2 =  d\B'_i dy_i  \wedge dx  \ , 
 $$
and $ \del_{[i} \B_{j]} =  \ha \ep_{ijk} \del_k H_5  ,  \ \ 
\del_{[i} \B'_{j]} =  \ha \ep_{ijk} \del_k H'_5.$
Dots in \ioi\ stand for higher-order terms coming from the $C_4$ 
background in  \wyve.
The leading terms in low-energy expansion are  
 \eqn\eey{
 I_3=\ha T_3 \int d^4\sigma 
   \big[ \  \kappa^{ab}
   ( H'_5\del_a y_i \del_b y_i + H_5 \del_a x_i \del_b x_i ) }
 $$ 
     +\  \ha  H_5   \kappa^{ac} \kappa^{bd } \F_{ab}\F_{cd} +  
     \F^{*ab}\hat C_{ab} + ... \big] \ .  
$$
 Assuming as above that 
the fields depend  only on the  first three world-volume 
 coordinates $\sigma_k \ (k=0,1,2)$
 and setting $A_3 \equiv \t$  we  find 
 \eqn\uoip{
 I_3=\ha T_3 \int d^4\sigma 
   \big[ \ H_5'(\vy) \del_k y_i \del_k y_i
+   H_5(\vx)  \del_k x_i \del_k x_i }  $$  
     +\ \ha  H_5 (\vx)   F_{kl} F_{kl} + 
      H'^{-1}_5 (\vy)   [\del_k \t  + \B_{i}(\vx)  \del_k x_i]^2
        +  \epsilon_{kls} F_{kl}\B'_{i}(\vy) \del_s y_i  + ... \big]  \ . 
$$
 Compared to the previous $5_{NS} \cap 5_{NS}$
example,  here the $5_{NS}$ monopole potential $\B$ couples to $F_{ab}$ 
 electrically while   the $5_{R}$ one  $\B'$ -- magnetically.
To decouple the $F_{kl}$ components 
of the field strength 
we introduce the `magnetic' variable $\td A_3\equiv \tt$
 as a Lagrange multiplier 
by adding the  term  $\epsilon_{kls} F_{kl}\del_s \tt$
which imposes the $dF=0$ condition 
(this is equivalent to 
performing the $d=3$ duality transformation $A_k \to \td A_3$).
Integrating out (or  redefining)  $F_{kl}$
we finish with (note that $\ep_{kls}\ep^{klr} = - \delta^r_s$, $\eta_{kl} = diag (-1,1,1)$) 
 \eqn\uoip{
 I_3=\ha T_3 \int d^4\sigma 
   \big[ \ H_5'(\vy) \del_k y_i \del_k y_i
+   H_5(\vx)  \del_k x_i \del_k x_i }
  $$   +\  
      H'^{-1}_5 (\vy)   [\del_k \t  + \B_{i}(\vx)  \del_k x_i]^2
        +   H^{-1}_5 (\vx)   [\del_k \tt  + \B'_{i}(\vy)  \del_k y_i]^2
  + ... \big]  \ . 
$$
As expected, the action is manifestly covariant under 
the  S-duality  transformation, i.e.
under $\vx \leftrightarrow \vy$, 
 $ H_5 \leftrightarrow H_5'$\  ($\B_i \leftrightarrow \B_i'$),  
 combined with world-volume 
duality  $\t \leftrightarrow \tt$.
This invariance  (extended to the full quantum level) 
 was related  in \hanw\ to a  mirror symmetry of   $N=4$ supersymmetric $d=3$
 gauge theories. 

The   moduli space metric corresponding to the 
$5_{NS} \cap 5_R$  configuration as measured by a 
classical 3-brane probe is thus 
 \eqn\meye{
  ds^2 =  H_5'(\vy) dy_i dy_i     + H^{-1}_5(\vx) 
[d \tt + \B'_{i} (\vy)  d y_i]^2  } 
$$ + \ H_5(\vx)  dx_i dx_i  +   H'^{-1}_5(\vy) 
[d \t + \B_{i} (\vx)  d x_i]^2 
  \  .  $$
In contrast to the 
metric  \yuy\  which appeared  in the $5_{NS} \cap 5_{NS}$
or $5_{R} \cap 5_{R}$ cases, this is a  non-trivial 
$D=8$ hyper-K\"ahler 
metric  \karl\ 
which does not factorise into a 
direct product of   independent 
 $D=4$ Euclidean  Taub-NUT metrics. 

One can repeat the above discussion for a $D=11$\   2-brane 
probe moving in the $5\cap 5 \botb 2$   background  \qqu.
The resulting action  for a 2-brane parallel to $(z,y_{11})$ is 
\eqn\mea{
 I_2=T_2 \int d^3\sigma \big[
    (-\det \hat  G )^{1/2}
    +{\textstyle  {1\ov 6} } \ep^{abc} \hat C_{abc}  } 
$$ = T_2 \int d^3\sigma
 \big( H_2\inv 
   \sqrt {\det [\delta^a_{b} + \kappa^{ac}
   ( H_2 H_5' \del_c y_n \del_b y_n + H_2 H_5\del_c x_m \del_b x_m ) ] } $$
$$ 
-  H_2\inv + \ha  \ep^{kl} [ B_{mn}(x)
 \del_k x^m \del_l x^n + B'_{mn}(y)  \del_k y^m \del_l y^n] \big)\ , $$
$$ \kappa_{ac} = diag(-1,1,H_5 H_5') \ , 
\ \   \ d B= *dH_5  \ , \ \ \ dB'= *dH'_5  \ ,   $$
where $a,b=0,1,2$, \  $k,l=0,1$.  Assuming that the fields do not depend on $\s_2=y_{11}$
and expanding in powers of derivatives 
we finish with an action which is a direct sum of the two 
5-brane  (super)conformal  2d models
\eqn\actio{
 I_2= \ha T_2 \int d^3\sigma \big[
H_5'(x)  \del_c y_n \del_b y_n  + \ep^{kl} B'_{mn}(y)  \del_k y^m \del_l y^n }
$$ 
+ \  H_5(x) \del_c x_m \del_b x_m + \ep^{kl} B_{mn}(x)  \del_k x^m \del_l x^n 
+ ... \big]\ . $$
This  is the expected result   as the $D=11$ $5\cap 5$
configuration  reduces to the two 5-brane NS-NS  background (cf. \qeqty). 

\newsec{Acknowledgements}
I am grateful to   M. Cveti\v c, J. Gauntlet, A.   Hanany,
I. Klebanov, J.  Maldacena and P. 
Townsend  for useful and stimulating 
discussions. 
This work was presented 
at  the Tel Aviv Workshop on Duality, January 9-12, 1997, and 
I would like to thank the organisers -- C. Sonnenschein and S. Yankielowicz
 for their kind hospitality. 
 I   
 acknowledge
 the support of PPARC and 
 the European
Commission TMR programme ERBFMRX-CT96-0045.
\vfill\eject
\listrefs
\end